# Monte Carlo Physarum Machine: Characteristics of Pattern Formation in Continuous Stochastic Transport Networks

Oskar Elek[1]    Joseph N. Burchett[1,2]    J. Xavier Prochaska[1,3]    Angus G. Forbes[1]

[1]University of California, Santa Cruz    [2]New Mexico State University    [3]Kavli IPMU

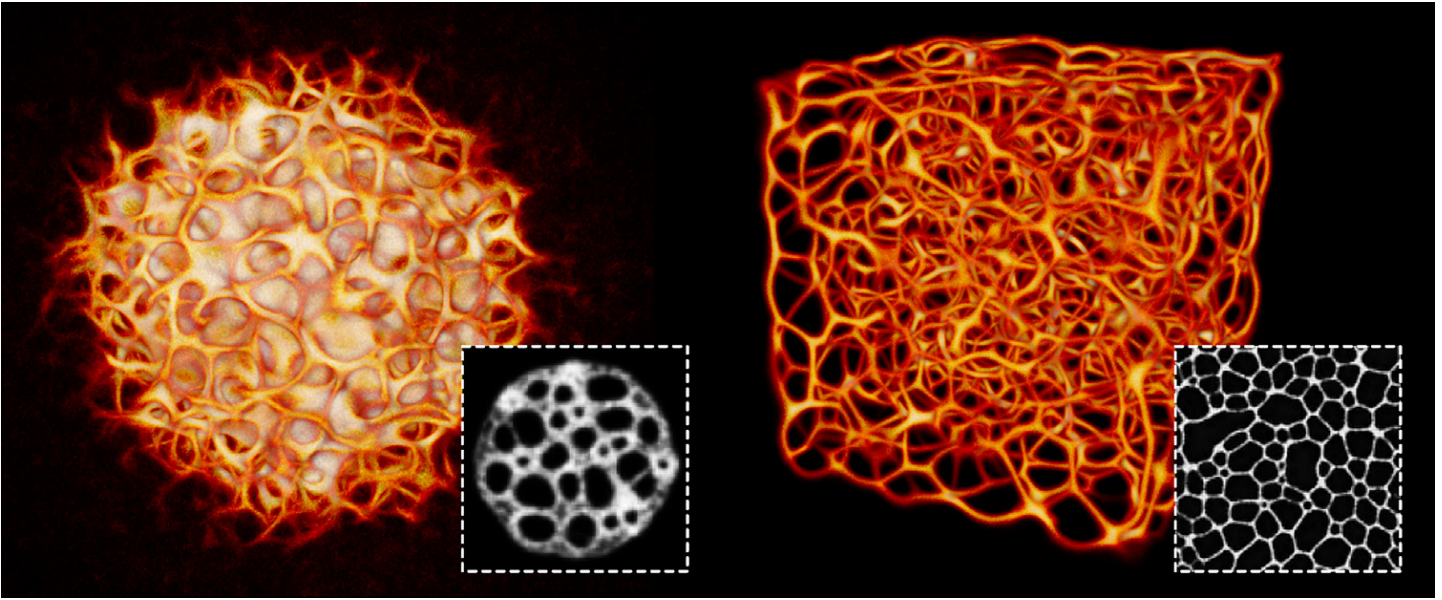

**Figure 1:** *Two examples of self-patterning results produced by the presented MCPM model. The grayscale insets show similar morphologies in 2D resulting from the model of Jones [2010], which MCPM generalizes to 3D by introducing a stochastic behavioral schema for the model agents.*

## Abstract

We present Monte Carlo Physarum Machine: a computational model suitable for reconstructing continuous transport networks from sparse 2D and 3D data. MCPM is a probabilistic generalization of Jones's 2010 agent-based model for simulating the growth of *Physarum polycephalum* slime mold. We compare MCPM to Jones's work on theoretical grounds, and describe a task-specific variant designed for reconstructing the large-scale distribution of gas and dark matter in the Universe known as the Cosmic web. To analyze the new model, we first explore MCPM's self-patterning behavior, showing a wide range of continuous network-like morphologies -- called "polyphorms" -- that the model produces from geometrically intuitive parameters. Applying MCPM to both simulated and observational cosmological datasets, we then evaluate its ability to produce consistent 3D density maps of the Cosmic web. Finally, we examine other possible tasks where MCPM could be useful, along with several examples of fitting to domain-specific data as proofs of concept.

## 1. Introduction

To understand a process, one must become it: the intelligence that evolved in nature therefore adapted itself to design pressures presented by various natural processes. To take an example that is central to this paper, the protist *Physarum polycephalum* 'slime mold' has been successfully used as an unconventional analog computer [Adamatzky2010,

Adamatzky2016]. As a 'thinking machine' and a creative pattern finder [Adamatzky2013], Physarum can solve hard spatial problems via its physical growth -- most prominently, finding optimal paths and transport networks connecting sets of data points where an underlying network structure is expected. The disadvantages, however, are significant: slow growth of the actual organism, difficulty of feeding it large inputs (e.g., more than a few dozen data points), and limited dimensionality. For these reasons, it is useful to instead use a virtual, simulated counterpart -- even if not as complex as its biological template -- as for many problems a simulation that qualitatively resembles Physarum's spatial behavior can be expressive enough [Jones2010, Jones2015].

The universe of problems involving interconnected network-like structures ever expands in the Information Age -- mostly due to communication and transportation networks (including road and utility networks, printed circuits, socio-economic networks, and the Internet). The importance of epidemiological network modeling has never been as apparent as in 2020/2021. And, in parallel to man-made networks, we continue to discover new natural phenomena with significant interconnected patterns: from molecular structures, through neuronal networks, circulatory/capillary systems, fungal and rhizomatic networks, all the way up to the network of dark matter filaments and intergalactic gas known as the Cosmic web. All these phenomena share an important feature: in one way or another they are built on the notion of optimal transport. Our initial concern in this line of work has been with the last case: reconstructing the filamentary structures of the cosmic matter distribution in a consistent and verifiable way.

To this end, we present **Monte Carlo Physarum Machine (MCPM)**: a simulated Physarum Machine that algorithmically mimics the organism's growth. Relying on Physarum's ability to closely approximate optimal transport networks, we feed to its virtual counterpart sparse point data as 3D attractors. MCPM's agents navigate these data to find most probable paths between them; the superimposed aggregate of their trajectories then forms the candidate network structure. We capture the totality of these trajectories as a density field in 3D space, giving rise to continuous network structures which we will refer to as "polyphorms" (see the examples in Figure 1). To demonstrate the practicality of the model, we apply MCPM to galaxy and dark matter halo datasets, interpreting the resulting continuous networks as proxies for the knots and filaments of the Cosmic web.

The first contribution of this paper is the full description of the MCPM model (Section 4). MCPM is a generalization of the seminal method described by Jones [2010], from which we draw much inspiration (Section 5). We expand Jones's work in several ways:

- All decisions in MCPM are **stochastic**. This makes the model significantly more customizable, as prior domain-specific knowledge can be encoded in probability density functions sampled by the model's agents.
- We extend Jones's model from **2D to 3D**, building on the stochastic formulation. We achieve this by redesigning the sampling stencil to work with only a binary sensing decision, rather than performing dense directional sampling as in Jones's proposed 3D extension [Jones2015]. Consequently, this choice also opens up the future possibility for higher-dimensional application of the method.
- We add a **new data modality** decoupled from the algorithm's signaling mechanism, to reconstruct the agent distribution. This modality, called *trace*, records the collective spatial history of agents' positions. After the model reaches equilibrium, the trace provides more accurate networks in conditions where the agent density by itself would be too low for getting robust reconstructions.

The second contribution is the case study of applying MCPM to datasets whose geometric complexity is higher than that of optimal transport networks but still resemble them to a large degree, especially with regard to their topology. Here we study the fitting process and reconstruction results using a task-specific variant of MCPM (Section 6), which we have previously applied in the context of cosmological modeling and visualization [Burchett2020, Simha2020, Elek2021]. We demonstrate how the probabilistic form of MCPM benefits the task of reconstructing the Cosmic web, the structural features of which are best described by a density field, rather than a network in the strict sense (Section 7). We posit that many real-world datasets share this property and offer several trials to support this (Section 8).

# 2. Background

## 2.1 Physarum Machines: real and virtual

Nature offers a wealth of inspirations for solving difficult computational problems, often referred to as *metaheuristics*. Natural selection itself can be framed as a constrained optimization process with selective pressures acting as design criteria. This is a fact that the tradition of evolutionary computation builds on [Goldberg1989, Hingston2008], as well as more directly biology-inspired optimization methods [Mirjalili2020].

Among the sources of inspirations, one organism clearly shines: *Physarum polycephalum*, commonly known as 'slime mold', has captured the attention of computational researchers over the past two decades (and half a century longer in biology [Bonner2009]). The growth patterns of the actual organism have been used to address many spatial problems: maze solving [Nakagaki2000], shortest path finding [Nakagaki2001], transportation design [Tero2010], the traveling salesman problem [Jones2014a], or Voronoi diagram estimation [Jones2015b], among others. A survey of this active area of research is given by [Sun2017]. Standing out is the extensive work of Andrew Adamatzky summarized in his books [Adamatzky2010, Adamatzky2016]. His team's recent work goes as far as proposing a Physarum-based hardware for solving suitable problems [Whiting2016].

The approach of using Physarum as an analog computer -- leading to the moniker **Physarum Machines** -- has been popular because of the organism's propensity to systematically explore its environment for food and shape itself into intricate networks to interconnect it. Food sources thus become straightforward proxies for input data, while different chemical and physical stimuli are available to further steer Physarum's growth [Adamatzky2010].

There are downsides too, however. Physarum grows slowly and is only suitable for small inputs (approximately up to dozens of distinct points). It's also surface-borne; and while heightfields are perfectly feasible [Evangelidis2017], truly 3D volumetric datasets are not. On the other hand, simulated slime mold -- a *Virtual Physarum Machine* -- can be designed to overcome these obstacles. VPMs usually simulate the organism *bottom-up*, obtaining Physarum's approximations from an iterated composition of simple, local rules.

In this work, we build on the simulation model of Jones [Jones2010]. While other computationally efficient VPM models have been proposed -- for instance, those based on cellular automata [Kalogeiton2015, Dourvas2019] and formal logic rules [Schumann2016] -- we adopt this agent-based model due to its ability to construct precise trajectories. Jones's model is related to agent-based modeling of crowds and flocks [Reynolds1987, Olfati-Saber2006], but uses a continuous representation of the agents' density for navigation [Jones2010, Jones2015]. On top of the successful application to 'classic' problems such as road network optimization [Adamatzky2017], this model is suitable for abstract and interactive tasks: B-spline fitting [Jones2014], convex and concave hull approximation [Jones2017], or robot navigation [Jones2015a]. In our case, it is the model's ability to *approximate optimal transport networks over a set of points* that made it an ideal base for us.

## 2.2 The search for the Cosmic web

The problem that originally motivated this work is the detection and reconstruction of the intergalactic medium (IGM), which forms the large-scale structure of the Universe (also known as the Cosmic web). This section briefly summarizes the existing work in this domain. For more details please refer to the comprehensive survey by Libeskind et al. [2018] as well as our prior work [Burchett2020, Simha2020].

Composed mainly of hot gas and plasma, the IGM is purported to trace the underlying dark matter scaffolding of the Cosmic web to contain a significant yet unobserved portion of the Universe's conventional matter: at least 40% estimated in total. This refers to the widely known *missing baryon problem* [Fukugita1998]. The IGM has long been thought to contain the unaccounted baryonic matter based on findings from large-scale cosmological simulations, such as the Millenium [Springel2005], EAGLE [Schaye2015] and Bolshoi-Planck [Klypin2016] simulations. Furthermore, the IGM hydrogen absorption lines frequently detected in spectra of distant quasars [Lanzetta1995] have long been assumed to trace the Cosmic web structure [Cen1994]. Recently, astronomers have used the dispersion measure of fast radio bursts to 'find' the missing baryons, verifying they reside in the IGM [Macquart2020]; however, important astrophysical problems remain, as the precise spatial distribution of the IGM is still heavily under-constrained.

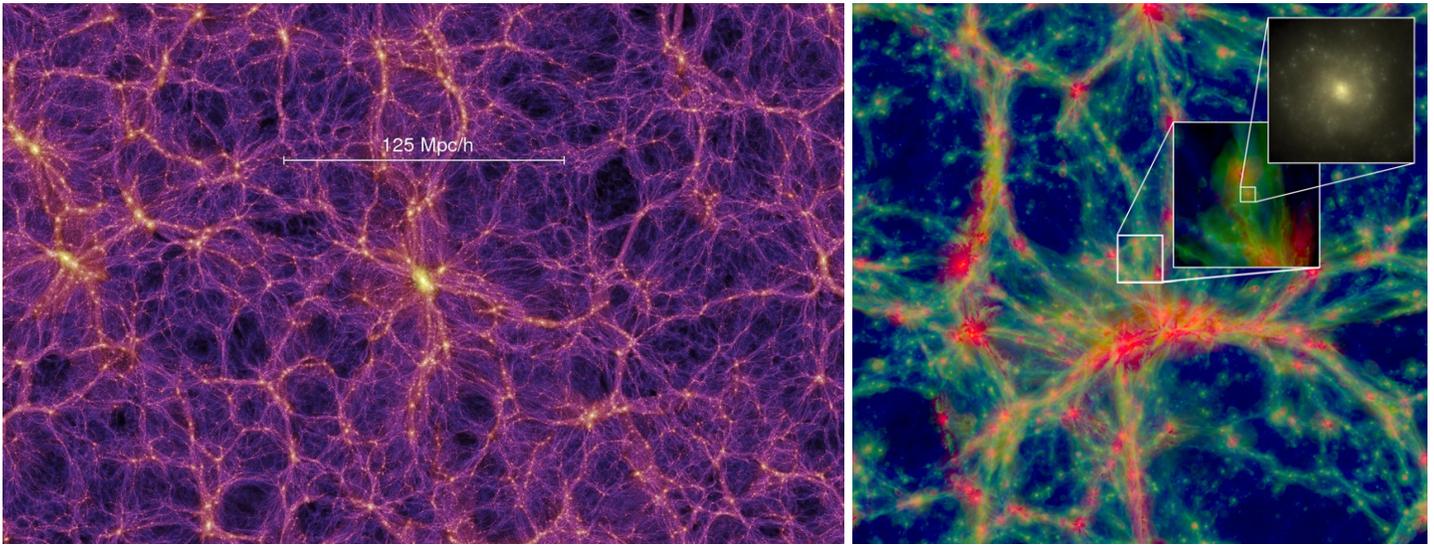

**Figure 2**: *Illustrating the scale of the Cosmic web structures: absolute in megaparsecs (left, Millennium simulation [Springel2005]) and relative in comparison to individual galaxies (right, EAGLE simulation [Schaye2015]).*

The large-scale simulations also unanimously agree on the expected distribution of the dark matter and thus the IGM: after billions of years, the combined effect of gravitational attraction and dark-energetic repulsion has shaped the universe as a vast network (Figure 2). This quasi-fractal structure [Scrimgeour2012], widely known as the **Cosmic web**, consists of *knots* (galaxies and their clusters) interconnected by a complex network of *filaments* and separated by vast cells of significantly under-dense space called *voids*. In addition, the entire structure is supported by a scaffolding of dark matter, according to theoretical predictions. These simulations themselves have been constructed to confirm the earlier findings of the Zeldovich framework [Zeldovich1970, Doroshkevich1970, Icke1973], which on theoretical grounds predicts the emergence of these very structures under the influence of the dominant force dynamics known in the Universe.

The pertinent question is: *how do we transfer the indirect knowledge about the Cosmic web to actually reconstructing its structure?* By definition, the IGM filaments would consist of diffuse gas that generally does not emit light at levels detectable by modern telescope instrumentation, making them extremely challenging to observe directly. On top of that, even the galaxies, which form within filaments and nodes and serve as the only luminous tracers of the Cosmic web, are observable only to a degree. For instance, with increasing *redshift* (i.e., distance from Earth), galaxies (particularly fainter ones) are more difficult to detect and may not be captured by astronomical surveys. We are therefore dealing with inherently incomplete, heterogeneous data.

The problem can be approached from multiple different angles. The most computationally amenable representations of network structures are based on *graphs* -- the field of combinatorial optimization [Cook1997, Papadimitriou1998] operates within this paradigm. Graphs can be regarded as topological structures [Porter2016, Wooodhouse2016] or even as a proxy for 3D geometry [Govyadinov2019]. Similarly in the astronomical context, both topological [Aragon-Calvo2010] and geometric [Tempel2014, Chen2016] approaches use graphs or partially connected graph-like structures as a core representation. Ultimately, most of the existing approaches have focused on cataloging and classification of the cosmic large scale structure [Libeskind2018]. To our knowledge, no available method is able to recover a complete 3D density map that is inherently driven by the underlying filamentary structure of possibly incomplete (especially observational) data. Given that the Cosmic web is a heterogeneous multi-scale structure without an intrinsic topology, having an accurate estimate for its *density distribution* is bound to benefit many use cases, including our prior work: better understanding the transition from the galactic to intergalactic environments [Burchett2020], characterizing the intergalactic medium towards singular astronomical phenomena [Simha2020], and potentially shedding new light on the missing baryon problem itself.

We approached this challenge from the perspective of finding a sound structural interpolator for the galaxy data, which on the scales where the Cosmic web exhibits distinct features (about 1 to 100 Megaparsecs) can be represented as a *weighted set of points* in 3D space. Optimally, the resulting model has to enable the following:

- detection of **distinct anisotropic structures** at least several Megaparsecs away from the galaxies, but ideally providing consistent density estimates in the entire region of interest;
- detection of **geometric features across several orders of magnitude** in terms of both spatial scale and density;
- **structural transfer**, allowing a calibration of the model on dense simulated data before deploying it on sparse observed data.

Per our findings, the presented biologically inspired swarm-based approach is able to cover all the above criteria. In this paper, we focus on the simulation and modeling perspective; more details towards the required astronomical and visualization tasks are provided in Elek et al. [2021].

## 3. Overview of Methodology

In its core, MCPM is a **dynamic computational model** defined by a set of priors: a set of probability density functions and their parameters obtained from domain-specific knowledge and/or fitting to training data. Once configured, the model can be fitted to input data: a weighted set of points in 2D or 3D (Figure 3, left). The result of this fitting is a continuous geometric structure (Figure 3, middle) which we interpret as a transport network over the input data. Rather than a graph, this geometry is represented as a *density field* stored as a densely sampled regular lattice (Figure 3, right).

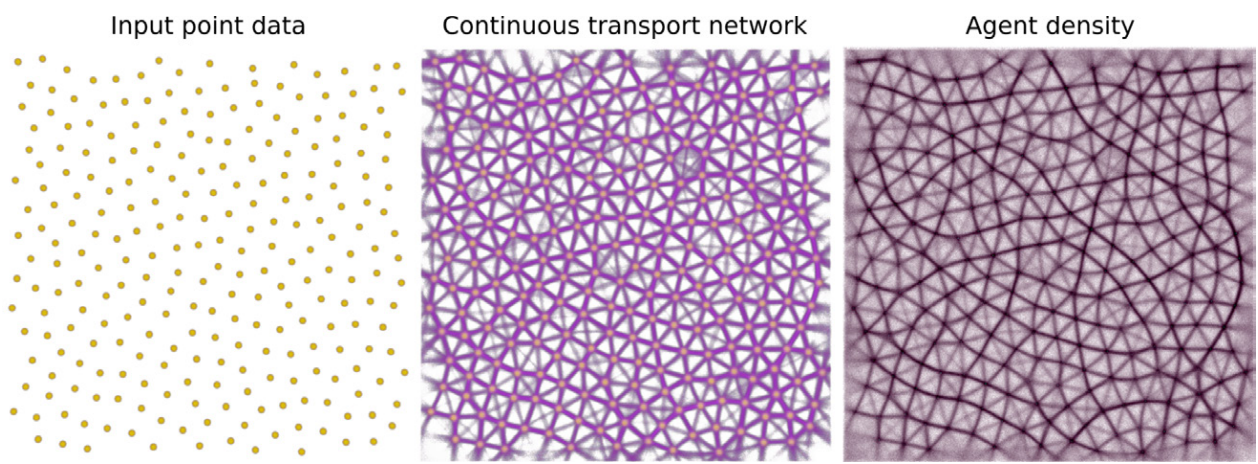

**Figure 3**: *Example of a simple continuous transport network reconstructed by MCPM from a set of uniformly distributed and equally weighted points in 2D. The network is akin to a continuous 'fuzzy' triangulation, with MCPM automatically localizing and connecting to the immediate neighbors of each input point.*

Section 4 defines the core components of MCPM on an abstract level, i.e. without specifying the prior probability density functions. In Section 5, we outline the differences between MCPM and the method of Jones [2010] and show that it is its natural generalization. Following that in Section 6, we detail the specific version of MCPM that we designed for the task of finding a meaningful Cosmic web map from observational data provided by astronomers: either galaxies or dark matter halos.

## 4. Monte Carlo Physarum Machine

MCPM is a hybrid model -- it has a discrete and a continuous component. **The discrete component** is an ensemble of particle-like agents that can freely navigate the simulation domain; these serve as a fragmented representation of the virtual organism. **The continuous component** is a 3D scalar lattice that represents the concentration of a marker that facilitates information exchange between the agents and the data. The model's behavior is based on a *feedback loop* between these two components, executed in two alternating steps: *propagation* and *relaxation* (see the attached pseudocode and diagrams in Fig. 4).

**(1) The propagation step** is executed in parallel for each of the agents, which are the model's device for exploring the simulation domain. Each agent's state is represented by a position and a movement direction, which are stochastically updated (Figure 4a-c) to navigate through the **deposit field** (referred to as 'trail' in [Jones2010], see Figure 4d, gray cells). The deposit field is stored as a 3D lattice of scalar values, representing the marker (in the biological context usually referred to as 'chemo-attractant') emitted by both the input data points as well as by the agents. The deposit effectively guides the agents towards the other agents as well as the data: the agents move with higher likelihood to the regions where deposit values are higher. In addition to the deposit, we also maintain a scalar **trace field** (Figure 4d, green cells) which records the agents' equilibrium spatial density, but does *not* participate in their guiding (detailed explanation of trace in Section 5).

```
def propagation_step( in params, inout agents[ ], inout deposit[ ][ ][ ], out trace[ ][ ][ ] ):
  in parallel for each agent i:
    ## treat data first
    if (agents[i] == "data")
        deposit[agents[i].pos] += agents[i].weight * params.data_deposit
        continue for

    ## agent sensing phase: panel (a)
    dir_s0 = agents[i].dir
    dir_s1 = sample(P_dir(params.sense_angle))
    dist_s = sample(P_dist(params.sense_distance))

    ## agent branching phase: panel (b)
    d0 = deposit[agents[i].pos + dist_s * dir_s0]
    d1 = deposit[agents[i].pos + dist_s * dir_s1]
    if (unit_random() < P_mut(d0, d1, params.sampling_exponent):
        dir_m = rotate_towards(dir_s0, dir_s1, params.move_angle)
    else:
        dir_m = dir_s0
    dist_m = sample(P_dist(params.move_distance))

    ## agent update phase: panel (c)
    agents[i].dir = dir_m
    agents[i].pos += dist_m * dir_m
    deposit[agents[i].pos] += params.agent_deposit
    trace[agents[i].pos] += dist_m / params.move_distance
```

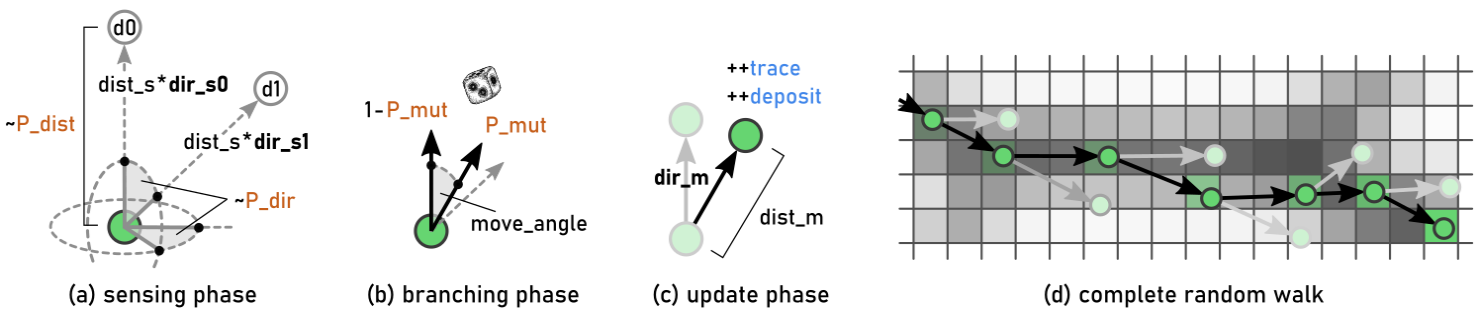

(a) sensing phase    (b) branching phase    (c) update phase    (d) complete random walk

**Figure 4:** Schematic depiction of the agent propagation behavior. By alternating the three stochastic sampling decisions (a-c), every agent describes a random walk (d). By choosing suitable sampling probabilities for the steps (a-c), the agents navigate the structure of the data-emitted deposit field (gray cells) and leave a trace marking their trajectory (green cells).

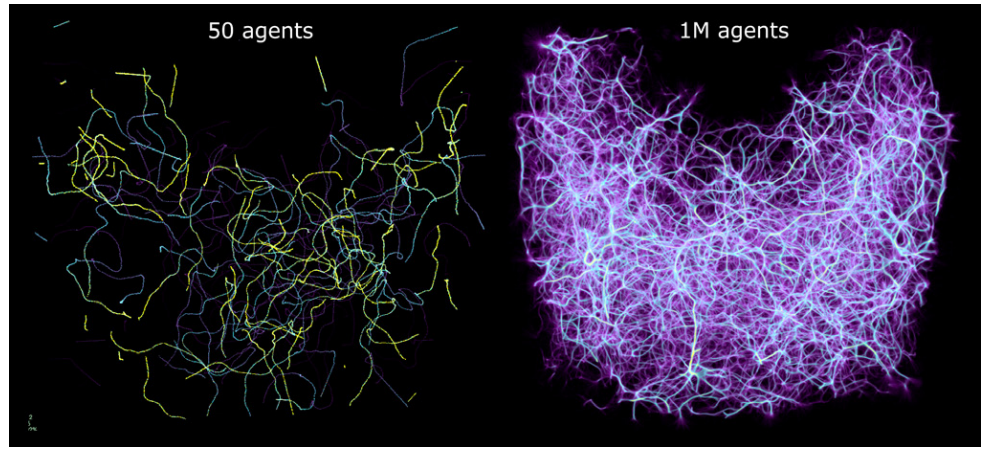

*Figure 5: Paths in 3D space drawn by the MCPM agents over 500 propagation steps (Section 4). While individual agents follow a seemingly random set of paths (left), a superposition of the entire swarm reveals an underlying structure present in the data (right). We store the resulting 3D structure as a density field called 'trace'.*

Successive propagation steps build up the agents' trajectories: *random walks* that follow the structure of the deposit field, and are recorded in the trace field. While each individual agent draws a seemingly chaotic path (Figure 5, left), a superposition of many agents averaged over a sufficient time window results in a smooth converged structure (Figure 5, right).

It is worth noting that the data points are also represented by agents, but of a special type: one that does not move, but merely emits deposit according to its weight ('mass'). We found this to be the most consistent way of handling the input data: the relative structural impact of the agents and the data now simply becomes a question of tuning the amount of deposit they each emit.

**(2) The relaxation step** ensures that the simulation eventually reaches an equilibrium. To this end, the *deposit field* is spatially *diffused* by a small isotropic kernel, and *attenuated*: that is, each cell is multiplied by a value < 1. The *trace field* is also attenuated but does not diffuse in order to preserve the geometric features in the agents' distribution, which the trace represents (please refer to Section 5 for further discussion on the trace field).

The simulation reaches its equilibrium when the amount of deposit and trace injected into the respective fields in the propagation step equals the amount removed by the attenuation in the relaxation step. With the parameters used in our experiments (Section 7), this usually takes hundreds of iterations.

```
def relaxation_step( in params, inout deposit[ ][ ][ ], out trace[ ][ ][ ] ):
    ## diffuse the deposit field
    in parallel for each voxel i in deposit:
        new_deposit = 0
        for each voxel j in neighborhood(i):
            new_deposit += deposit[i] / normalized_distance(i,j)
        synchronize_threads()
        deposit[i] = new_deposit

    ## attenuate deposit and trace
    in parallel for each voxel i, j in deposit, trace:
        deposit[i] *= params.attenuation
        trace[j] *= params.attenuation
```

## 4.1 Probabilistic sampling

The core of MCPM is the agent propagation defined in terms of three **probability distributions**, the configuration of which can be tuned in runtime. These are:

- $P_{dir}$ for sampling the agents' directional decisions during the sensing and movement phases; defined on the unit sphere relatively to the current direction of propagation.
- $P_{dist}$ for sampling the agents' distance decisions during the sensing and movement phases; defined in positive real numbers along the current propagating direction.
- $P_{mut}$ for making the binary 'mutation' decision whether the agent should remain on its current course, or branch into the newly sampled direction according to $P_{dir}$.

By specifying these three distributions one defines a particular *instance of MCPM* -- we discuss this further in the following Section 5. For our Cosmic web detection task, we detail our choices of these three distributions in Section 6.

Now, it is good to take a step back and understand *why* such a model produces spatial networks and why these align with the input data. Figure 6 supports this discussion. The key is that the agents are navigated towards large 'pools' of deposit -- this is ensured by defining $P_{mut}$ such that it preferentially selects those directions where higher deposit values were sensed. Agents are therefore attracted towards data in their vicinity (thus interconnecting them), as well as other nearby agents (thus reinforcing existing pathways). The shape characteristics of the network (curvature of the paths, size and acuity of the features, as well as connectivity patterns) are further dependent on the combined influence of $P_{dir}$ and $P_{dist}$. In particular, $P_{dir}$ needs to be strongly forward-oriented, which ensures momentum preservation and prevents the agents from getting stuck in local pools of built-up deposit.

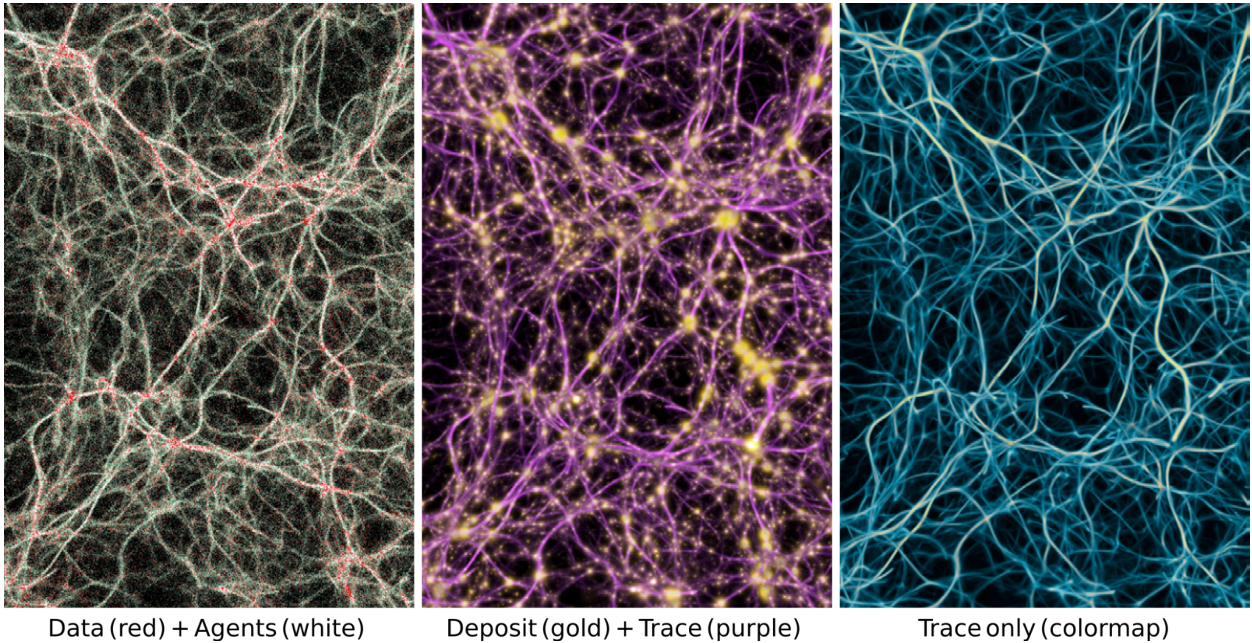

Data (red) + Agents (white)　　Deposit (gold) + Trace (purple)　　Trace only (colormap)

**Figure 6:** *Summary of all the data modalities that MCPM operates with. Left: individual agents (white particles) flowing among the data points (red particles). Middle: pools of deposit emitted by the data (gold) interconnected by the trace field (pink). Right: the resulting filamentary structure contained in the trace field, representing the transport network.*

In summary, MCPM solves tasks by acting as a *structural interpolator* for input data. When adequately configured and parametrized, it can find natural transport networks that connect neighboring points as a human observer would intuitively do -- even in the topologically richer 3D space (Section 7). These networks are represented by the converged trace, i.e., by a continuous density field rather than an explicit graph.

## 5. Relation to Max-PM

Our main motivation for extending the 'Max-PM' model of Jones [2010] was the question of *parametrization*. While Max-PM is configurable enough to reproduce the morphology of Physarum Polycephalum and beyond, the agents respond to different deposit concentrations in always the same way: *follow the direction of maximum deposit concentration*. Hence the alias "Max-PM", PM standing for Physarum Machine.

At any rate, this behavior -- especially after our initial experiments in 3D -- lead to overly condensed pathways and only moderately connected networks (more in Section 7). While captivating and resembling a 3D Physarum, the fits that Max-PM produced did not match our target cosmological data well. Another observation was the necessary increase of

*sampling effort*: Max-PM in 2D uses 1+2 samples for directional sensing, our extension to 3D needs at minimum 1+8 directions (determined empirically). This increases the sampling effort considerably as well as the number of access operations to the deposit lattice.

These considerations lead to us targeting the agents' directional sampling, specifically modifying it to use stochastic Monte Carlo sampling. In the remainder of this section we discuss our specific extensions of Max-PM in detail.

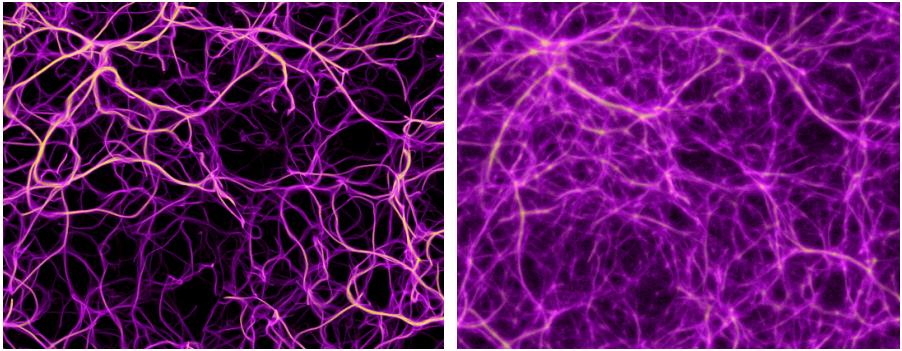

*Figure 7:* A small segment of a transport network grown on the same data using Max-PM (left) and MCPM (right). Max-PM yields a clean, well defined structure, which however does not consistently cover all the input data, in contrast to the MCPM variant (32% data points missed by Max-PM, compared to 0.072% missed by MCPM).

## 5.1 Stochastic sampling of agent trajectories

The way in which Max-PM treats **agent navigation** is suitable for the original context: the Physarum agents use chemotaxis to move around, following the chemoattractant concentration gradient (represented by the deposit field). In ours and arguably many other scenarios -- where the main concern is recovering a network structure that fits target data -- we need a more complex behavior that can be configured based on *available knowledge about* the data distribution. The main concern is that in any given geometric configuration of point data a multitude of feasible pathways is available to meaningfully interconnect them.

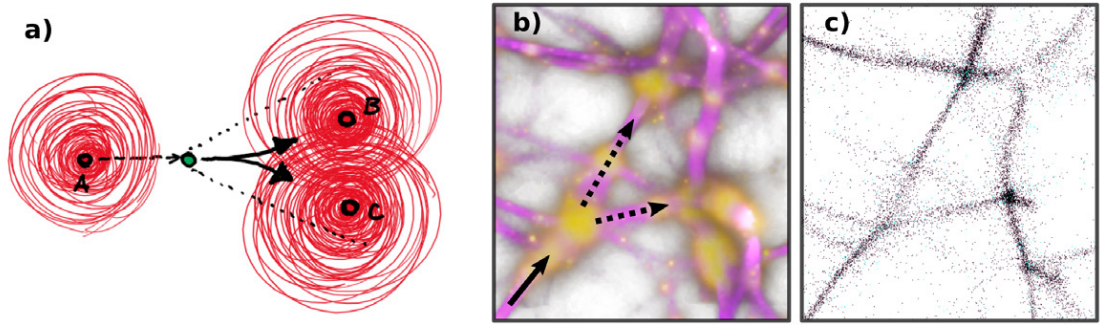

*Figure 8:* Simple configuration where the agents arriving from data point A need to split evenly between points B and C (a). In an actual reconstruction scenario, this corresponds to a bundle of agents splitting about evenly and branching out into two separate filaments (b,c) which afterwards merge with the flow of the other adjacent agents.

As an example, consider the elementary configuration in Figure 8a: we want the resulting network to branch out when connecting the data point A with B and C, which means that the agent traveling from A has to choose steering towards B or C with roughly the same probability. This ensures that the aggregate distribution of all agents passing through this location is going to have a branched shape. In Figure 8b-c we see an actual such configuration: agents arriving from the bottom-left branch split in two 'streams' and fluently transition to the more complex region in the right.

This behavior is represented by the *mutation decision* (Section 4) encoded by the discrete probability $P_{mut}$ -- that is, whether the agent should remain on its current course, or branch out in the direction generated in the sensing step.

To provide additional flexibility in representing paths with different scales and curvatures, we modified the agents' spatial decisions to behave probabilistically as well, rather than using constant steps and angles for the agents' movement [Jones2010]. This behavior is defined by the continuous probability density functions $P_{dir}$ (2D distribution on a sphere) and $P_{dist}$ (1D distribution on a half-line).

We can now demonstrate how Max-PM can be framed as a *special case of MCPM*.

- Define the mutation probability $P_{mut}$ as $d_1^\infty / (d_0^\infty + d_1^\infty)$, where $d_0$ is the deposit value ahead of the agent and $d_1$ is the deposit in the mutated direction. Such definition effectively means that the larger of the deposits will always be selected for the agent to follow.
- Define $P_{dir}$ and $P_{dist}$ using Dirac delta distributions with the angular and distance parameters as offsets for the peaks.

In Section 6 we discuss the $P_{dir}$, $P_{dist}$ and $P_{mut}$ that define the particular variant of MCPM we employed for reconstructing the Cosmic web structure.

## 5.2 Extension from 2D to 3D

Enabled by the notion of the mutation probability $P_{mut}$ we can now simplify the *sampling stencil* -- that is, how many directions need to be sampled during the agent's sensing stage. We observed that in Max-PM the network complexity (connectivity) depends on the number of directional samples, and that this becomes even more pronounced in the topologically richer 3D space.

Given that the directional navigation in MCPM is controlled by $P_{dir}$ and $P_{mut}$, we can reduce the necessary number of samples to two: one in the forward direction, and one in the mutation direction. Any direction where $P_{dir}$ is nonzero can potentially be sampled, and even directions with low deposit have a chance to be selected (subject to the specific definition of $P_{mut}$). This has two advantages: the savings in directional sampling can be reinvested into increasing the resolution of the agent's trajectory, and extensions to higher than three dimensions are now possible without increasing the sampling effort.

The fact that high-dimensional sampling and integration problems can be effectively solved with binary sampling stencils is well established in the *Monte Carlo simulation* community. Both direct [Kajiya1986, Veach1997, Kalos2008] and Markov-chain Monte Carlo [Metropolis1953, Hastings1970, Veach1997a] avoid the *curse of dimensionality* this way when constructing random walks. Moreover, this strategy can even be the most efficient one, as documented, for instance, by the ubiquity of path tracing methods for simulating light transport in the visual effects industry [Christensen2016, Fong2017]. Indeed, our method draws significant inspiration from these works.

## 5.3 Aggregation of agent trajectories

Finally -- in addition to the deposit field -- we define an additional data modality in MCPM: the **trace field**, or simply "trace". Just like the deposit, the trace is stored in a 3D lattice with the same resolution (Figure 4d and Figure 6, middle). Trace is computed as a superposition of all locations visited by each agent within a certain time window -- it is an equilibrium spatial distribution of agent trajectories. Please refer to the pseudocode in Section 4 for how the trace is constructed on the algorithmic level -- formally, we can define it as a probability density $P_T$ over positions **x** in 3D space:

$$P_T(x) = n \cdot E[N(x)]$$

where **N** is the number of agents in the cell adjacent to **x**, and **n** is a normalization constant. The expected value of $N(x)$ is the Monte Carlo estimate calculated as a moving average, as a function of the attenuation parameter (cf. pseudocode for the relaxation step in Section 4).

Rather than using the deposit to recover the reconstructed network (as in Jones's method [2010, 2015]) we use the trace for this purpose. This has multiple advantages:

1. Resolution: in contrast with the deposit, the trace field is not subject to spatial diffusion -- structural details are therefore preserved (Figure 9).
2. Specificity: recovering the network from the trace becomes much easier than from the deposit, as in the latter, the data term has by design a much higher impact than the agent term.
3. Interpretability: while the exact mathematical properties of the trace in relation to $P_{mut}$, $P_{dir}$ and $P_{dist}$ are yet unknown, they are nevertheless in a direct relationship. One way to interpret the trace is the *total probability* of the agents' distribution marginalized over all possible geometric configurations that can occur in a given dataset.

Another way of interpreting the trace is that it provides a continuous estimate of the geometric and topological structure outlined by the input data. This is related to the use of such a structure by Jones and Saeed [2007], referred to as 'trail' in their work. The application of trail here was to substitute for standard image-processing kernels, as used for denoising or contrast enhancement. Unlike their trail, the trace lives in a different frame of reference than the input data: the data are sparse and unordered points, while the trace is a function densely sampled in space and continuous. Other than that, these data structures are conceptually similar.

To succinctly describe the trace in text, we establish the term **"polyphorm"**. We will be using this term to refer to the particular continuous geometries represented by the trace field, as well as a general label of the concept of a continuous transport network, i.e., a network *implicitly* defined by the trace density field (as opposed to a network explicitly represented by a graph or other means).

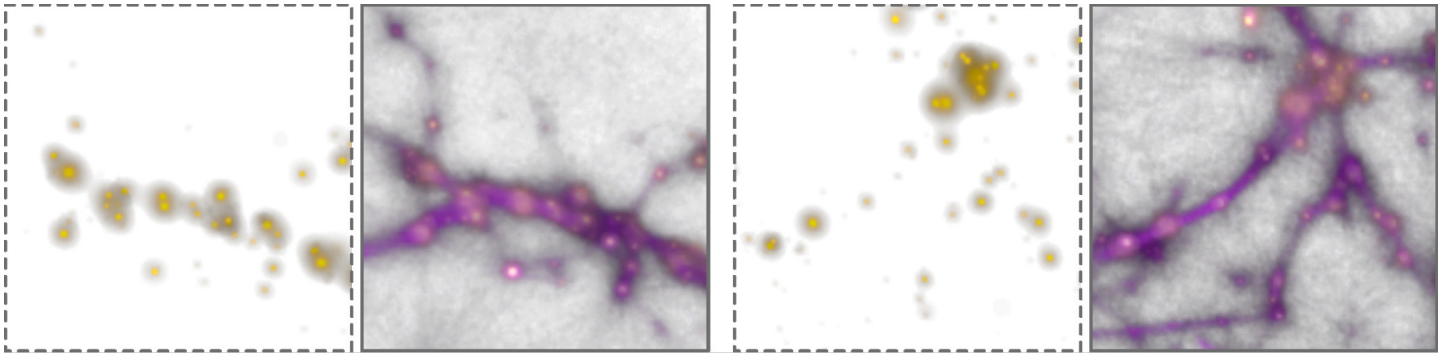

**Figure 9**: *Comparing the distribution of data-emitted deposit (gold) and the reconstructed trace field (purple) in two sample configurations. While the deposit field is sparse and diffused, the trace is continuous and sharp, and thus serves as a better estimate of the transport network.*

## 5.4 Fully continuous representation

Our last design choice is to fully decouple the representation of agents and the underlying discrete data structures. In Max-PM, *only one agent* can be present in each cell of the deposit field at any given time. This is suitable for simulating the actual organism, as each agent represents a small fragment of the body and therefore occupies a certain volume of space.

In our case, this behavior is no longer desired: our agents represent an abstract spatial distribution with possibly several orders of magnitude in dynamic range. For instance, cosmic overdensities between $10^{-3}$ and $10^2$ are important in the astrophysical inquiry. We therefore allow agents to *move freely* without enforcing any limit on their number in each grid cell. This has additional benefits: it decreases the amount of necessary bookeeping, and decouples the reconstructed network structure from the data structures' resolution (as demonstrated in Section 7).

## 6. MCPM for Cosmic Web Mapping

In Sections 4 and 5 we described the components of MCPM and provided rationale behind our design. Now we specify the probabilities $P_{dir}$, $P_{dist}$ and $P_{mut}$ which together define the variant of MCPM used in our previous work [Burchett2020, Simha2020, Elek2021] for the Cosmic web reconstruction task. We likewise use it in the experiments throughout Section 7 and most of the paper.

#### Directional distribution $P_{dir}$

The main criterion for $P_{dir}$ is that the agents should maintain a certain *momentum* -- otherwise, if allowed to turn too rapidly, they would invariably get stuck in local maxima of the deposit field. The simplest choice here is uniform distribution over the forward-facing cone of directions. This distribution is parametrized by a single scalar parameter: the opening angle of the cone. Routines for efficiently drawing samples from this distribution can be found for instance in [Watson1982].

Other viable choices of $P_{dir}$ include the spherical Gaussian, the Von Mises-Fisher distribution, the Henyey-Greenstein distribution [Henyey1941], and their truncated versions.

When configuring $P_{dir}$, the rule of thumb we used in our fitting confirms the findings of Jones [2010]: the sensing angle should be larger than the mutation (i.e. turning) angle. We typically use twice the value for sensing than for turning. The variation of these and other parameters is further explored in Section 7.

#### Distance distribution $P_{dist}$

Since the Cosmic web is built up from filaments consisting mostly of hot gas, we use the *Maxwell-Boltzmann distribution* for sampling the agents' sensing and movement distances. This statistical distribution describes the velocity dispersion of particles in idealized gas and plasma [Mandl1988], which makes it suitable for our use case and potentially for modeling other physical systems. Efficient sampling from this distribution is described e.g. by Hernandez [2017].

Other good candidates for $P_{dist}$ include the exponential distribution (e.g. to model bacterial motion dynamics [Li2008]), log-normal (for particles under Brownian motion), Poisson, Gamma, and other distributions defined in $\mathbb{R}+$. Convex combinations of multiple modes are also possible, if different distinct scales of features are expected in the modeled system. The weight of the tail in $P_{dist}$ determines the range of feature scales MCPM will be able to reconstruct: in the simplest case of constant sensing and movement distances, the model will be sensitive to only one characteristic feature size.

When generating the sensing and movement distances for an agent, we use the same seed to sample $P_{dist}$ -- this is to avoid the cases when the movement distance would exceed the sensing distance. Even more so than in the directional sampling, here we find that the sensing distance should be significantly larger than the movement distance: a good starting point is a difference of about an order of magnitude.

#### Mutation probability $P_{mut}$

In contrast to distributions $P_{dir}$ and $P_{dist}$, this is a binary probability that an agent deviates (branches away) from its current course. Its definition needs to ensure that agents *predominantly* steer towards higher concentration of the deposit. (This, in our experience, is the sufficient condition for the simulation to equilibrate in a configuration that follows the data.) If this were the opposite case, the agents would actually be *repulsed* by the data, as well as from each other. With reference to the pseudocode in Section 4 we define $P_{mut}$ as

$$P_{mut}(d_0, d_1, s) = d_1^s / (d_0^s + d_1^s)$$

where $d_0$ is the deposit in the forward direction, $d_1$ is the deposit in the mutated direction, and $s >= 0$ is the sampling exponent. We examine the impact of the sampling exponent (and other parameters) on the resulting network geometry in Section 7.

An alternative definition of $P_{mut}$ could assume that the agents move in a participating medium with density proportional to the deposit, in which case the agent dynamics would be subject to the Eikonal equation and $P_{mut}$ could be derived from the Beer-Lambert law. This is not our case however, as collisions between the intergalactic medium particles do not have appreciable impact on the Cosmic web geometry.

It is worth noting that by increasing the values of the sampling exponent we approach the limit case discussed in Section 5, i.e., MCPM becomes identical with Max-PM for $s \to \infty$.

# 7. Experiments

Our open-source implementation of MCPM called *Polyphorm* (`github.com/CreativeCodingLab/Polyphorm`) is written in C++ and uses DirectX with GPU compute shaders for executing the parallel propagation and relaxation steps, as well as facilitating an interactive visualization.

Our **GPU implementation** was developed on an NVIDIA TitanX, and is capable of fitting 10 million agents over $1024^3$ deposit/trace grids at roughly 150-200 ms per model iteration, with an additional 30-100 ms needed for the visualization. Including the agents and rendering buffers, no more than 6 GB GPU memory is consumed at that resolution, using float16 precision to store both the *deposit* and *trace* grids. In all our experiments the model converges in fewer than 700 iterations (about 1-3 minutes of real time). It is thanks to the massive parallelism of modern GPUs, combined with the high locality of memory accesses in MCPM, that we are able to simulate 10M or more agents at interactive rates. This is essential for obtaining noise-free simulation results. Further performance data are available in Elek et al. [2021].

The **performance** scales close to linear with the number of voxels, and sub-linearly with the number of agents: 100M agents runs at 300 ms *ceteris paribus*, that is 10x more agents at only 2x the slowdown (compared to 10M agents). Using more agents has minimal impact on the model's spatial details, but it does increase *effective resolution* by proportionally reducing the Monte Carlo noise levels.

All of the model's **parameters** (specific values are provided in the following subsections, with reference to the pseudocode in Section 4) can be modified at any point during the simulation, which allows for easy tuning of the resulting 3D polyphorm structures. Edits typically take several seconds to manifest, both visually and in the value of the loss function (in Section 7.3 we discuss the details of our employed loss function).

We rely on **cosmological datasets** to conduct the practical part of the evaluation, in which we focus on fitting the model. These include the Bolshoi-Planck large scale cosmological simulation [Klypin2016] at zero redshift (i.e., corresponding to the present-day Universe) with 840k dark matter halos extracted by the Rockstar code [Behroozi2012]. These are simulated data with a-priori known ground-truth Cosmic web structure, which is why we use that to *calibrate* the MCPM parameters. Our target dataset for reconstructing the Cosmic web structure comprises 37.6k galaxies (their locations and masses) from the Sloan Digital Sky Survey (SDSS) catalog [Alam2015], spanning redshifts between 0.0138 and 0.0318. For more details about the astronomical evaluation of our results, please refer to [Burchett2020, Simha2020]. Here we will be concerned with the behavior of MCPM and the geometry generated by it.

For the **visualizations**, we use direct volume rendering [Beyer2015] (for field data) and particle rendering (for the agents). Sometimes, in addition to projections of the full 3D data, we isolate slices and cross-sections where better insight is necessary. Full exposition of our visualization methodology is provided in Elek et al. [2021].

All the following experiments use the MCPM variant defined in Section 6, unless stated otherwise.

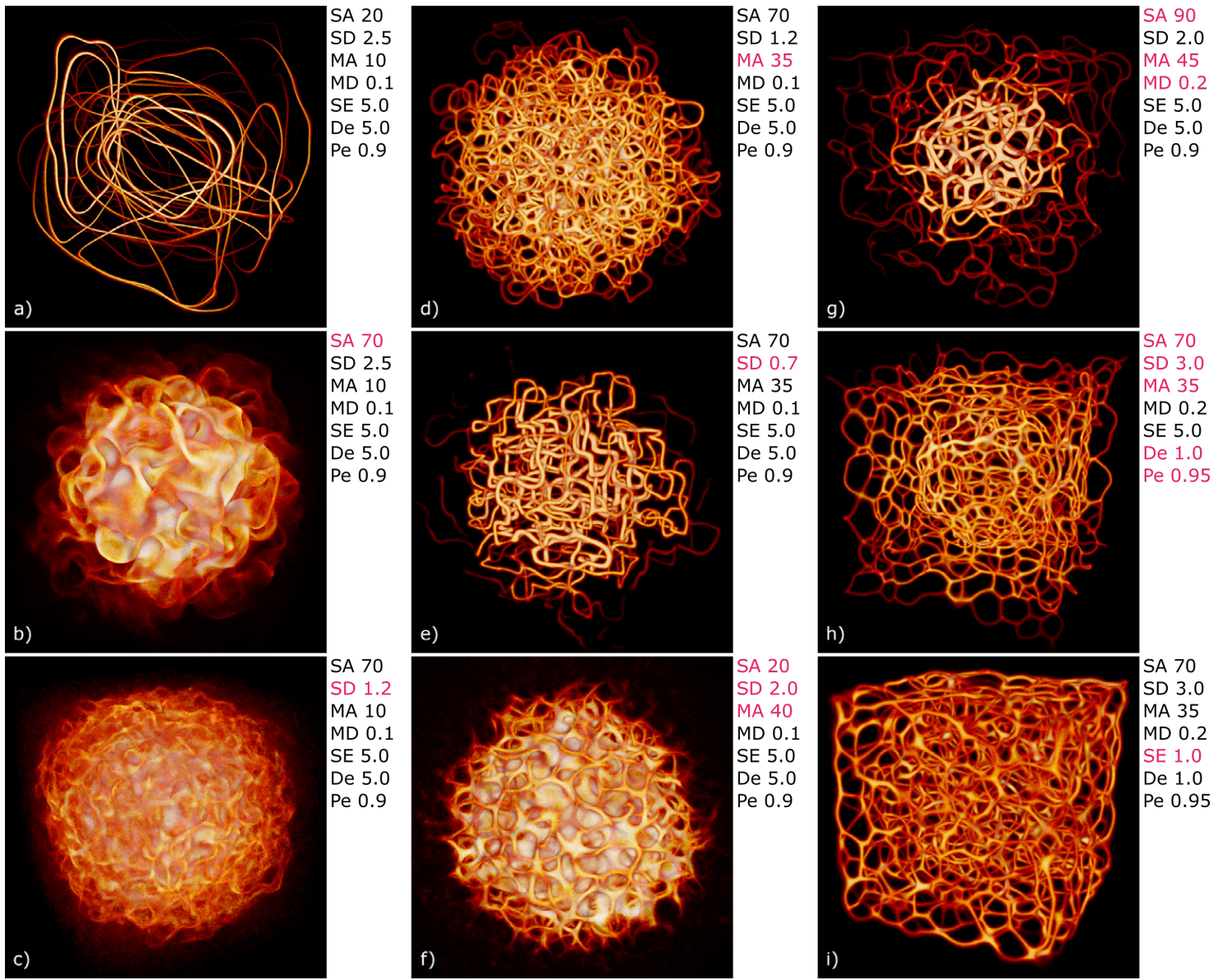

**Figure 10:** *Interesting polyphorms the model generates without any pre-patterning cues, including features like tubular filaments, loops, fuzzy membranes, thin fibers, branches, and undulating pulses. Some of these morphologies closely resemble those in Jones [2010] (such as d, f, h), others are similar but topologically connected in novel ways (e, g, i), and others still are completely novel (a, b, c). In all of these results, a weak attracting force towards the center of the domain has been applied to the agents, in order to prevent crowding along the edges which can sometimes happen. Starting at (a), every subsequent change of MCPM parameters is highlighted. The parameters (see pseudocode in Section 4) are Sensing angle (SA), Sensing distance (SD), Move angle (MA), Move distance (MD), Sampling exponent (SE), agent Deposit (De) and field Persistence (Pe). The size of the simulation domain is 100 units in each dimension.*

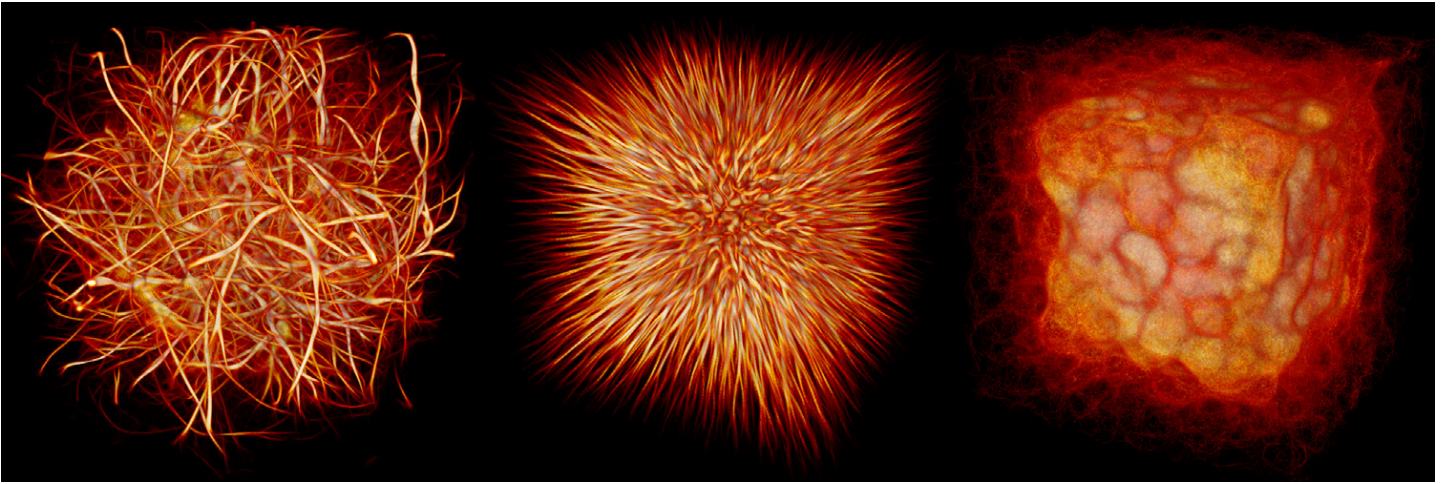

**Figure 11**: *Unstable polyphorms generated by MCPM, representing phase transitions between different configurations. A focused examination of transient phenomena generated by another Physarum simulation has been done by Jenson and Kuksenok [2020], studying the interaction with complex artistic and didactic systems.*

## 7.1 Self-patterning

Even though we primarily care about the data-driven mode of operation, we start by looking at polyphorms produced by the model without any prior stimuli. In the following examples we work with a $540^3$ simulation grid uniformly filled with 2M agents at the start of the simulation.

In line with Jones's findings [Jones2010] we observe that the free patterns created by MCPM are *stable in the qualitative sense*. That means that for any given parametrization the generated features will be comparable, but rarely static: the generated polyphorms are in constant flux. In the experiment in Figure 10 we explore a number of configurations that lead to distinctive results. This experiment was conducted in a single session, changing one or multiple MCPM parameters at a time and waiting for the simulation to sufficiently converge after resetting the agents to random initial positions.

In addition to these meta-stable patterns, a plethora of *transient* ones is accessible. These occur when the model's configuration changes significantly, as a sort of dynamic phase change. Figure 11 shows a few illustrative examples of serendipitous trips through the parameter space. Given that these forms are dependent on the momentary state of the model before the phase change, and are unique to both the original and modified parametrization, there is currently no way to enumerate them systematically.

Even though all the parameters impact the polyphorm, the *most significant determinants* of shape are the Sensing angle, Sensing distance, and the respective movement parameters. These would end up to be the parameters we adjusted the most frequently during fitting. In the experiment in Figure 12 we therefore focus on exploring these fitting dimensions. Curiously, we observe correlation between the stability of the fit and how interconnected the network structure is: the polyphorms in the upper right part of the figure were much more persistent in time than those in the left.

Another important design dimension is to consider different behaviors of the agents. In particular, many natural systems contain repulsion as another important shaping force. Encoding this type of behavior in MCPM simply means to reverse the sampling decision given by $P_{mut}$ (see pseudocode in Section 4). Consequently, the agents will navigate towards lower deposit concentrations, i.e., locations not occupied by other agents or data points. More complex repulsion behaviors could also be encoded through specialized redesigns of $P_{dir}$ and $P_{mut}$.

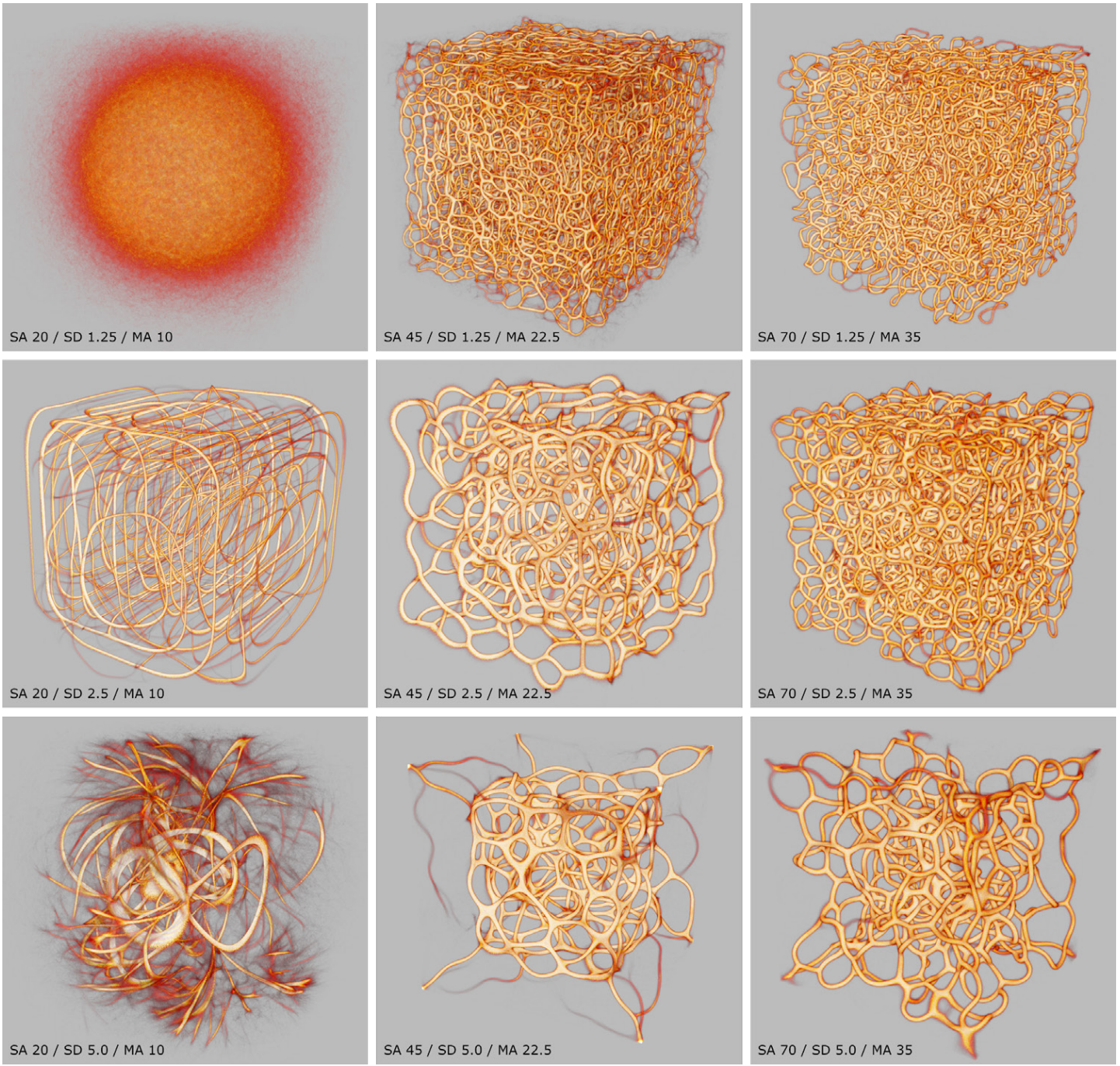

**Figure 12:** *Varying the Sensing angle (SA) horizontally and the Sensing distance (SD) vertically. The Move angle is fixed at half the SA and Move distance at a constant 0.1, Sampling exponent 3.0, Agent deposit 2.0 and Persistence 0.9. The domain size is 100 units.*

Figure 13 documents three MCPM runs in $1024^3$ simulation grids and 10M agents, comparing the original attractive agent behavior, its repulsive inversion, and a 50/50 mix of the two behaviors. In line with the previous experiments, the attractive behavior leads to a network-like polyphorm that, after forming, remains relatively stable and slowly condenses to thicker and sparser structure. On the other hand, the repulsive behavior forms cellular patterns around the sites where they were initialized. These patterns are only transient however: as the agents continue repelling each other, the pattern breaks down and dissipates. Finally the mixed behavior -- where each agent randomly decides to be attracted or repelled by deposit -- leads to a structure that is a blend of both: sites form cells, neighboring sites merge together, and the resulting interconnected structure slowly diffuses. Figure 13 maps the temporal progression of each regime, and additionally shows a 3D view of the entire structure at a point when its features are most visible.

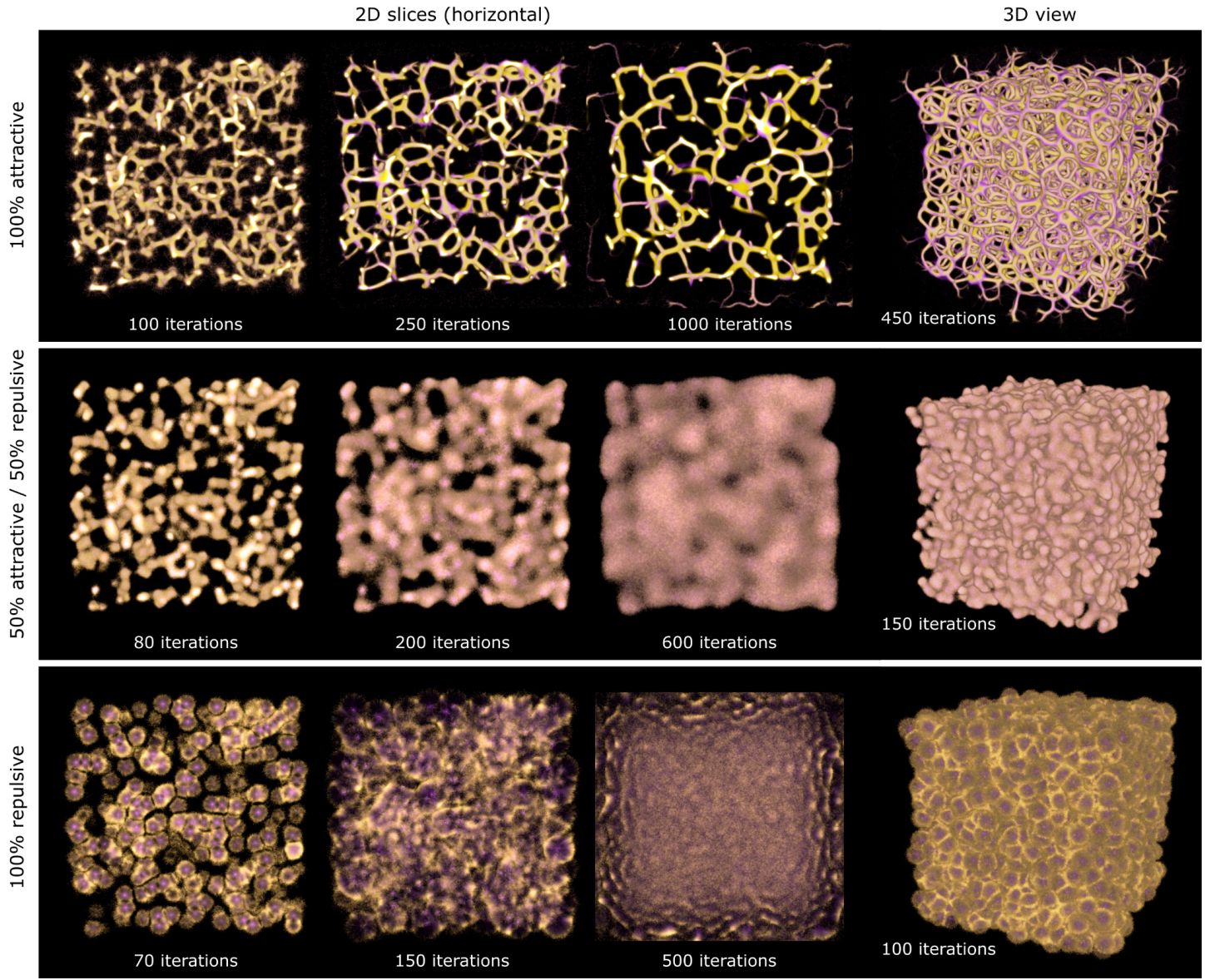

**Figure 13:** *Polyphorms generated by an ensemble of attractive agents (i.e., the base MCPM, top row), repulsive agents (bottom row) and a 50/50 combination of the two behaviors (middle row). The agents were initialized at random sites in a domain 100 units in size, with no data present. The visualization shows a superposition of both the trace (purple) and deposit (yellow). All the experiments were generated using the following parameters: Sensing angle 60 deg, Move angle 30 deg, Sensing distance 3.0, Move distance 0.05, Agent deposit 10.0, Persistence 0.95, Sampling exponent 5.0.*

## 7.2 Uniform stimuli

We now start introducing data in *homogeneous configurations*. We show that MCPM produces meaningful networks over points with different uniform distributions in 3D: in a regular lattice (Figure 14), distributed randomly (Figure 15), and following a low-discrepancy 'blue noise' distribution [Ahmed2016] (Figure 16).

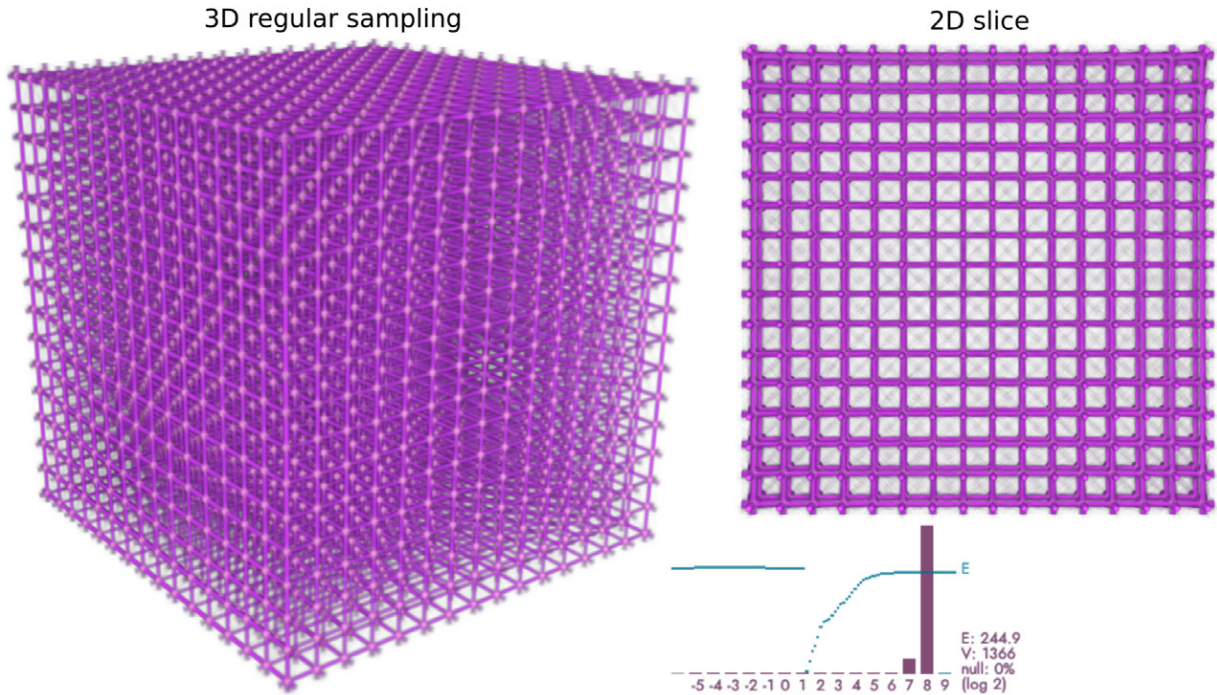

**Figure 14:** Regular lattice of $16^3$ = 4k points, simulated using deposit/trace grids with $720^3$ voxels (vox), 10M agents, Sensing distance (SD) 50 vox, Move distance (MD) 0.65 vox, Sensing angle (SA) 20 deg, Move angle (MA) 10 deg, Sampling exponent (SE) 5. The plot shows the log-2 histogram of trace readouts at the data points (bordeaux) and the fit energy (cyan).

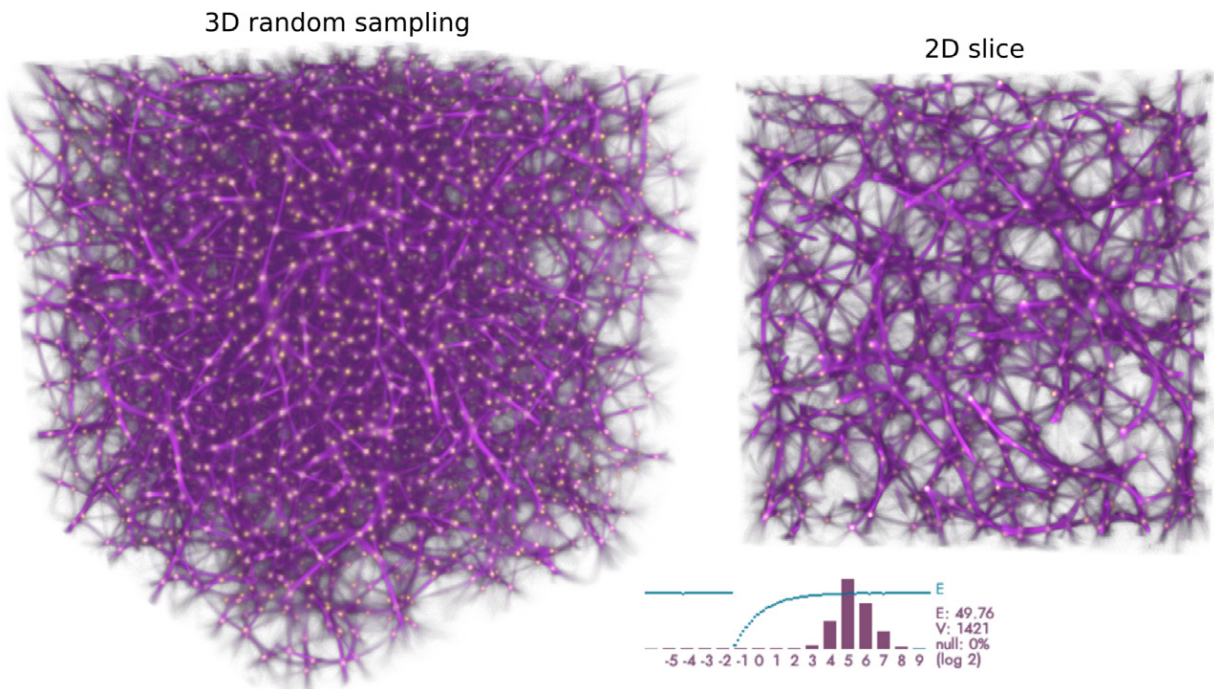

**Figure 15:** Volume with 4k uniformly-random distributed data points. Simulated using deposit/trace grids with $720^3$ voxels (vox), 10M agents, Sensing distance (SD) 50 vox, Move distance (MD) 0.65 vox, Sensing angle (SA) 20 deg, Move angle (MA) 10 deg, Sampling exponent (SE) 5. The plot shows the log-2 histogram of trace readouts at the data points (bordeaux) and the fit energy (cyan).

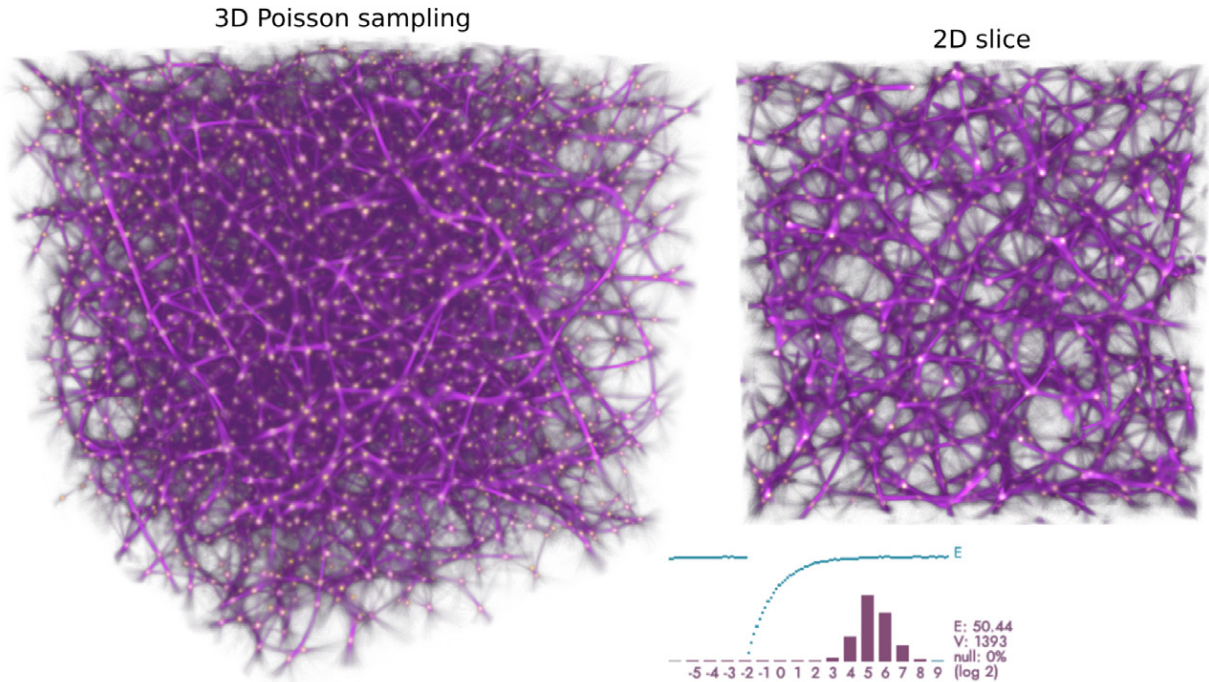

**Figure 16**: *Volume with 4k uniformly distributed points with a low-discrepancy Poisson-like distribution. Simulated using deposit/trace grids with 720³ voxels (vox), 10M agents, Sensing distance (SD) 50 vox, Move distance (MD) 0.65 vox, Sensing angle (SA) 20 deg, Move angle (MA) 10 deg, Sampling exponent (SE) 5. The plot shows the log-2 histogram of trace readouts at the data points (bordeaux) and the fit energy (cyan).*

In a regular lattice (Figure 14) MCPM robustly finds straight connections between points and due to the multiscale movement of the agents (cf. Section 6) also the diagonals of the sub-boxes. The model can be tweaked to prefer the diagonals simply by increasing SD to about 70--80 vox.

The fitting takes about 300 iterations of the model to converge (see the discontinuity in the energy plot in Figure 14, bottom, where we restarted the fitting process). Remarkably all the data points receive *nearly identical density* of the agents, as shown in the histogram.

Randomly distributed points (Figure 15) have a *tendency to cluster*, leading to a spurious reinforcement of some of the fitted network connections. As a result, fitting from uniformly initialized agents takes somewhat longer, about 500 iterations. The density histogram at data points is close to Gaussian, with a slight positive skew, and variance proportional to the density fluctuation of the data.

In Figure 16, using a low-discrepancy Poisson-like distribution of the data [Ahmed2016], we obtain perhaps the most organic of homogeneous patterns. Since the low-discrepancy points emulate the distribution of cells in biological tissues (and similar such phenomena), the resulting fit is resembling a natural scaffolding, not unlike bone marrow (also cf. Section 8.3). The structure also bears resemblance to rhizomatic systems (such as tree roots and mycelial networks).

The fitting takes about 500 iterations and results in a density distribution not unlike the random case, but with a slightly stronger positive skew. Tweaking the values of SE and then trying different distance sampling distributions would impact the acuity of the fitted network, and would expand or narrow the resulting distribution histogram.

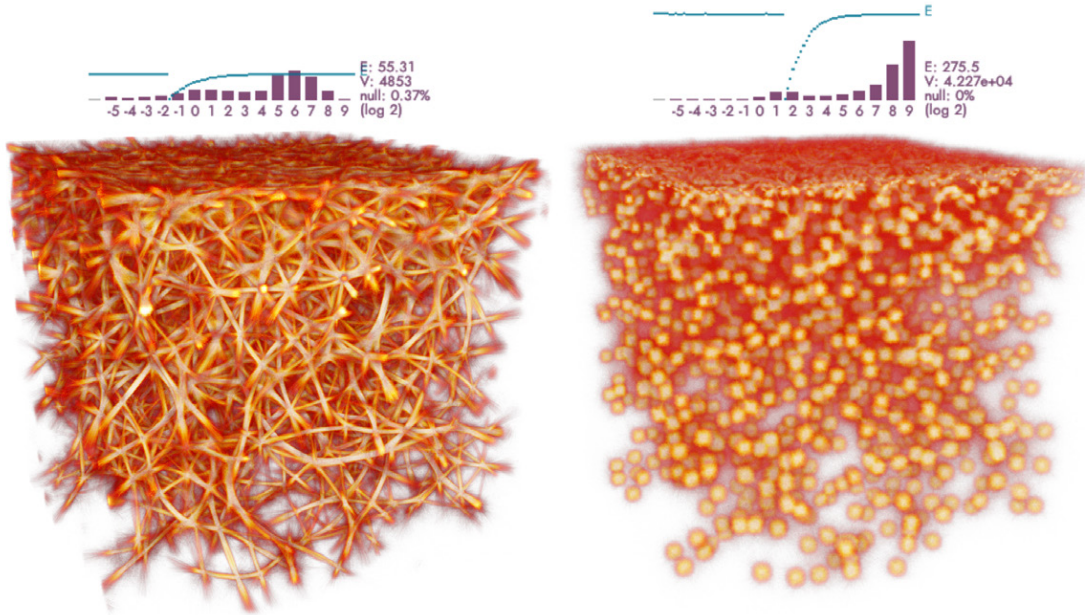

**Figure 17**: Two fits to 4k evenly distributed points but with density biased towards the top of the domain. Both fits use identical fitting parameters (akin to Figures 14-16), with one difference: in the right, the agents use 7x smaller sensing distance compared to the left case. This change alone causes a collapse of a healthy, interconnected network into a set of isolated 'islands'. Please note that the histograms in this figure plot the overall trace distribution in the whole volume (rather than trace at the data points only as in all other figures).

## 7.3 Fit energy (loss function)

MCPM solves the problem of reconstructing a volumetric transport network over a set of discrete points. The network should include all input points, perhaps except for outliers. We also want the weight of the input points to influence the strength of the network connections, i.e., to attract more agents.

The energy function we have used for fitting encapsulates all of the above considerations. To compute the energy $I$ of the current MCPM fit, we compute the mean of the trace values summed over the input data point locations, normalized by the data points' weights:

$$I(trace) = \frac{1}{data.count} \sum_{d \in data} \frac{trace[d.pos]}{d.weight}$$

This design has one shortcoming: *collapsed fits* can occur, where all the agents are attracted to the nearest data point (Figure 17, right). Compared to a 'healthy' transport network (Figure 17, left) the collapsed fit lacks connections between neighbors, rather resulting in a disconnected set of 'blobs'. Since there is no established definition of connectedness in this context, we leave this aspect for visual evaluation and use the above defined energy function to only explore the fitting parameter space *locally*. The interactive visualization in *Polyphorm* and the ability to inspect the fit by slicing and different rendering modes was therefore essential to obtaining good reconstruction results [Elek2021].

## 7.4 Bolshoi-Planck data (simulated)

Our first evaluation of an actual reconstruction scenario is based on the Bolshoi-Planck (B-P) cosmological simulation at redshift zero [Klypin2016] with dark matter halos extracted by the Rockstar method [Behroozi2012]. *Halos* are considered to be significant clusters of dark matter and are the formation sites of galaxies; the most massive halos are typically located at the intersections (knots) of filaments. Importantly, at the scales of our interest the halos are sufficiently represented by their 3D position and mass. The full dataset contains 16M halos, most of which have extremely low mass and are not significant in determining the overall filamentary structures. Therefore, we remove all halos with mass $M_{halo} < 5 \times 10^{10}\ M_{sun}$, leaving approximately 840k of the significant ones. These halos span a cubic volume approximately 170 Mpc on a side.

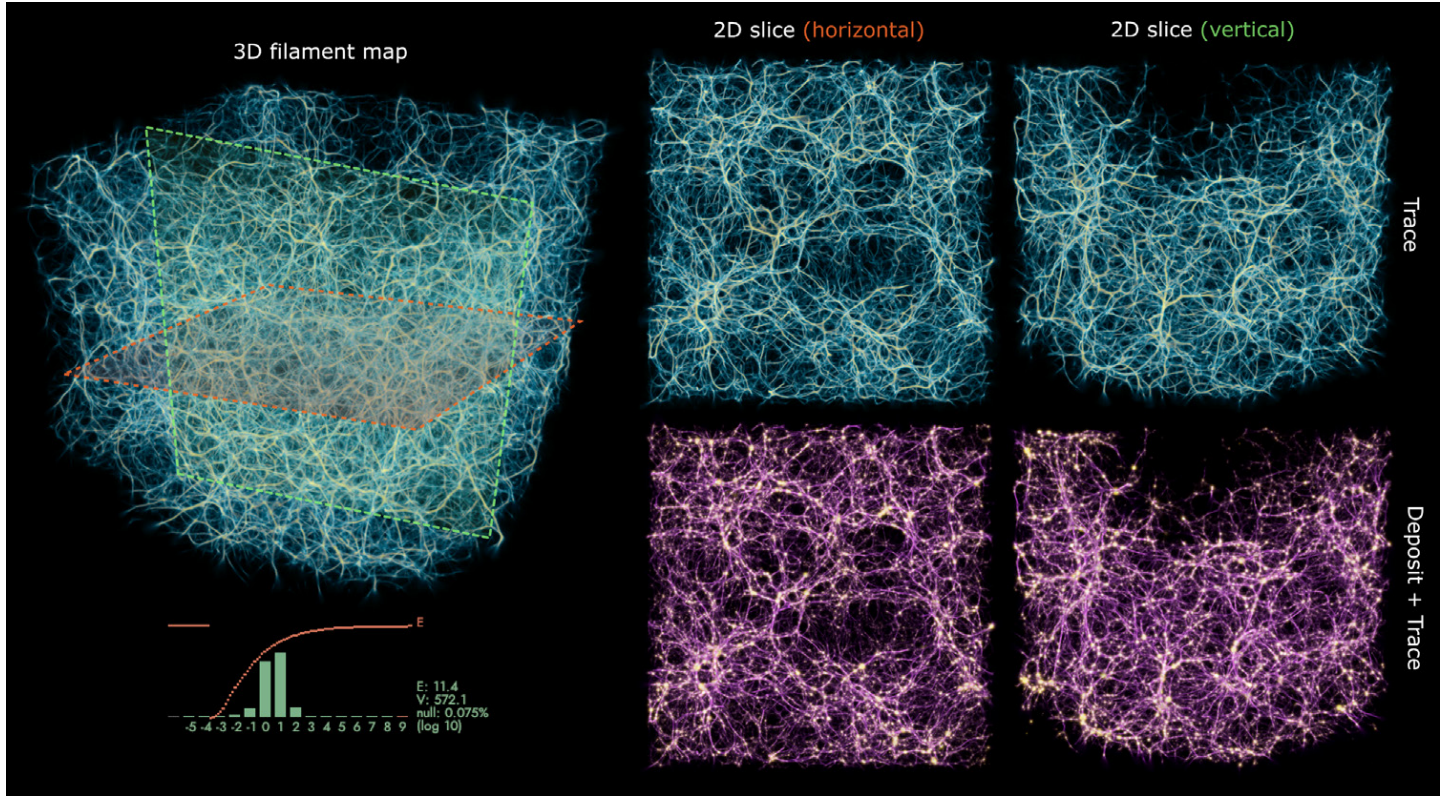

**Figure 18**: MCPM fit to 840k halos from the Bolshoi-Planck dataset, simulated using deposit/trace grids with $1024^3$ voxels (vox), 10M agents, Sensing distance (SD) 2.5 mpc / 13.7 vox, Move distance (MD) 0.1 mpc / 0.57 vox, Sensing angle (SA) 20 deg, Move angle (MA) 10 deg, Sampling exponent (SE) 3.0. The simulation domain was 187 Mpc across (with 170 Mpc spanned by the data). The halos are restricted to a spherical 'shell' centered around an imaginary point of origin (this in order to emulate the redshifts where our SDSS dataset is defined, see Section 7.5).

Because dark matter is gravitationally coupled to conventional, baryonic matter (on large scales), it follows that their distributions will mostly have the same structure. Also, since the B-P dataset is simulated, it does not suffer from key limitations of observed data: namely, data points missing due to occlusions or other observational issues. This makes the B-P data an ideal *fitting prior*, and as such we use them to **calibrate** the MCPM model's parameters to further use with the observed Sloan Digital Sky Survey (SDSS) data (Section 7.5).

Figure 18 shows the MCPM fit to B-P data. We obtain a detailed polyphorm with features on *multiple scales*: in line with theoretical predictions, we observe characteristic loops and filaments on scales from a single Mpc to about 100 Mpc. The *quasi-fractal* nature in the data is also revealed -- we observe self-similar knots and voids (lacunae) at different scales.

Fitting to the B-P data takes about 900 iterations and (similarly to the uniformly distributed point sets) results in a near-Gaussian density distribution at the halo locations, as seen in the histogram in Figure 18. The fits match the underlying data extremely well: more than 99.9% of the halos are covered by the reconstructed network, and as detailed in Burchett et al. [2020] we find a strong correlation between the MCPM trace and the ground-truth particle density of the B-P dataset. This was crucial to establish the mapping between the MCPM trace density and the *Cosmic overdensity*, to yield interpretable astronomical results. Further numerical evaluation of this correlation is provided in

Section 7.10. The progression of the fitting and the formation of the obtained geometric features is documented in Figure 19.

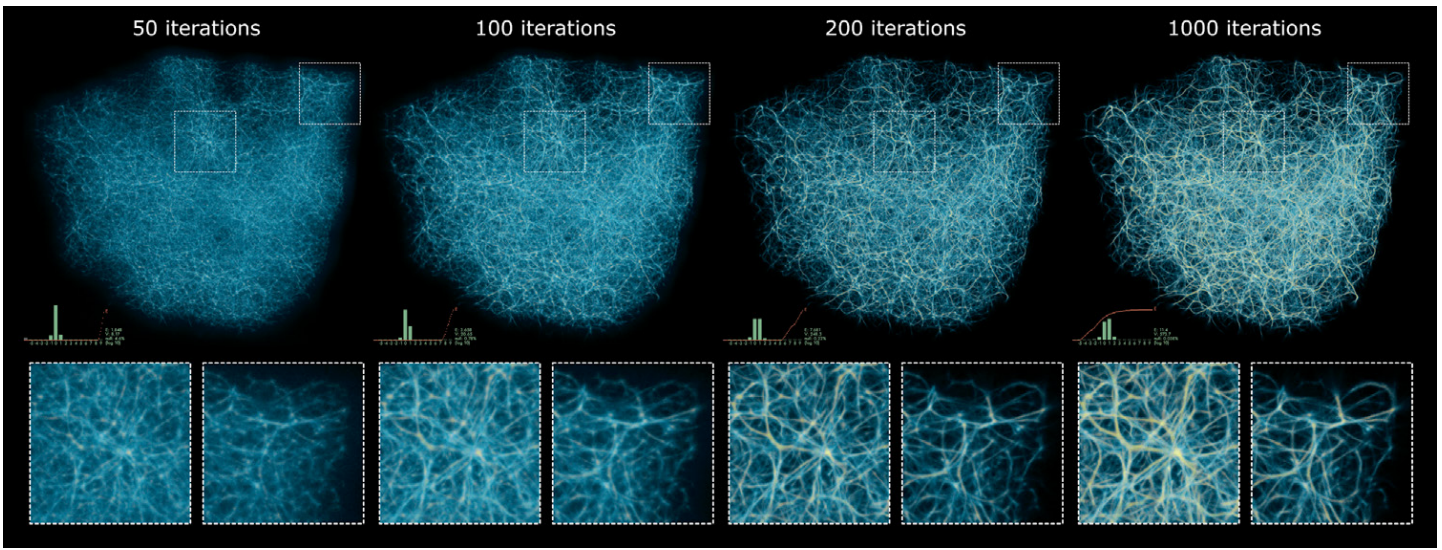

**Figure 19**: *Convergence behavior of MCPM when fitting to the 840k B-P halos. Due to the large size of this dataset the fitting takes longer on average; however, already after a fraction of the full fitting time we obtain a decent estimate of the eventual converged polyphorm (as seen in the insets below), allowing for early adjustments of MCPM's parameters.*

## 7.5 Sloan Digital Sky Survey data (observed)

Having verified in Section 7.4 that MCPM is capable of fitting simulated structures arising from the current cosmological theory, we now move to the actual observed data. The target here is to reconstruct the Cosmic web spanning 37.6k spectroscopically observed galaxies extracted from the Sloan Digital Sky Survey (SDSS) catalog [Alam2015]. The selected galaxies are within the redshift range [0.0138, 0.0318], equivalent to the distances between 187 and 437 million light years from Earth, which on Cosmic scales is relatively close to home. Just as with the Bolshoi-Planck data, the galaxies at these scales are sufficiently represented as points in 3D space (derived from the celestial coordinates of each galaxy and its corresponding redshift-derived distance) with weights proportional to the galactic masses. After the spatial transformation of galaxy coordinates, the galaxies are contained within a volume with 280 Mpc in its largest dimension.

The main challenge faced here is the *incompleteness of the data*: some galaxies elude observation either due to their low apparent brightness, or being too closely projected on the sky to other bright galaxies. Whatever the case, we need MCPM to be robust under these conditions. The most important precaution is to prevent the model from 'detecting' filaments where there are not sufficient observations to support them. To this end, we disable the agent-emitted deposit to make sure that all detected structures are due to actual data.

Figure 20 shows the MCPM fit to the SDSS data. To obtain this result, we use the same model parameters as for fitting the Bolshoi-Planck data, thus facilitating a prior-informed feature transfer. In line with that, the fitted polyphorm ends up with structurally similar features -- in spite of fitting to an order-of-magnitude fewer points, we observe filaments, knots and voids on the expected scales. The density distribution of the network likewise exhibits a clearly near-Gaussian profile, as well as similar convergence characteristics (Figure 21).

The MCPM fit also reveals an artifact in the data: the so-called *fingers of god*, elongated structures pointing towards Earth (visible in the vertical 2D slices in Figure 20, right). These result from biased conversions of redshift to line-of-sight distance in the observations; compensating for them is therefore subject to corrections applied to the data itself -- for more details please refer to the discussion in Burchett et al. [2020]. Currently we do not apply any such corrections.

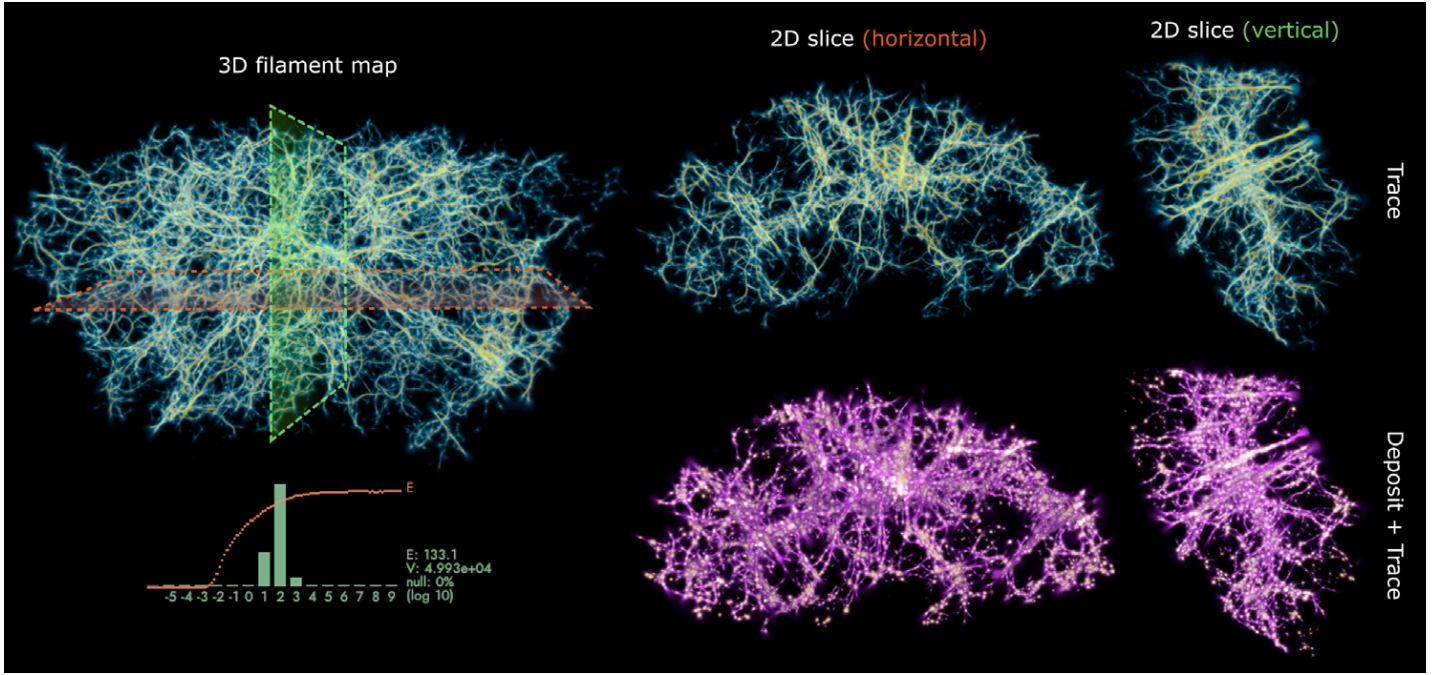

**Figure 20**: *MCPM fit to the 37.6k galaxies in the SDSS catalog, simulated using deposit/trace grids with 1024x584x560 voxels, 10M agents, and otherwise identical parameters as in the Bolshoi-Planck case. The simulation domain was 310 Mpc across the largest dimension.*

A key **validation** of this fit is based on correlating the reconstructed filament network with spectroscopic sightlines from distant *quasi-stellar objects* ('quasars'). This is the main result in Burchett et al. [2020], in which we found a strong positive correlation between neutral hydrogen absorption signatures and the MCPM-derived density in low-to-medium density regimes, and a significant *anti-correlation* in the high-density regime. The latter can be attributed to the lack of neutral hydrogen in the close proximity of galaxies, as hydrodynamical shocks and other processes associated with galaxy formation ionizes the hydrogen gas and thus suppressing its ability to absorb the light from quasars.

Crucially, what enabled the validation of the MCPM-generated map is the *continuous nature* of the fit. Rather than obtaining a binary polyphorm detecting the presence of agents, the probabilistic nature of MCPM yields a scalar trace field expressing the time-averaged density of the agents. We interpret this trace simultaneously as a density and a probability that the detected features are significant; as described above, we establish a monotonic mapping between the MCPM trace and the cosmic overdensity, which is a directly interpretable quantity in the astronomical context.

At this point, in reference to the requirements formulated in Section 2, we can conclude that the model qualifies for the designated task of **reconstructing the Cosmic web** map from observed data:

- MCPM naturally synthesizes anisotropic filamentary features, and, when configured correctly, finds a cohesive transport network over the input data.
- MCPM outputs a continuous density map with features across multiple levels of scale and intensity.
- MCPM can be easily pre-trained on prior data and transferred to target data through the optimized values of its (hyper)-parameters.

In the remainder of this section, we perform additional experiments to probe the model for its specific behaviors and robustness under degradation of the input data. We choose the Bolshoi-Planck dataset for this evaluation due to its higher point density.

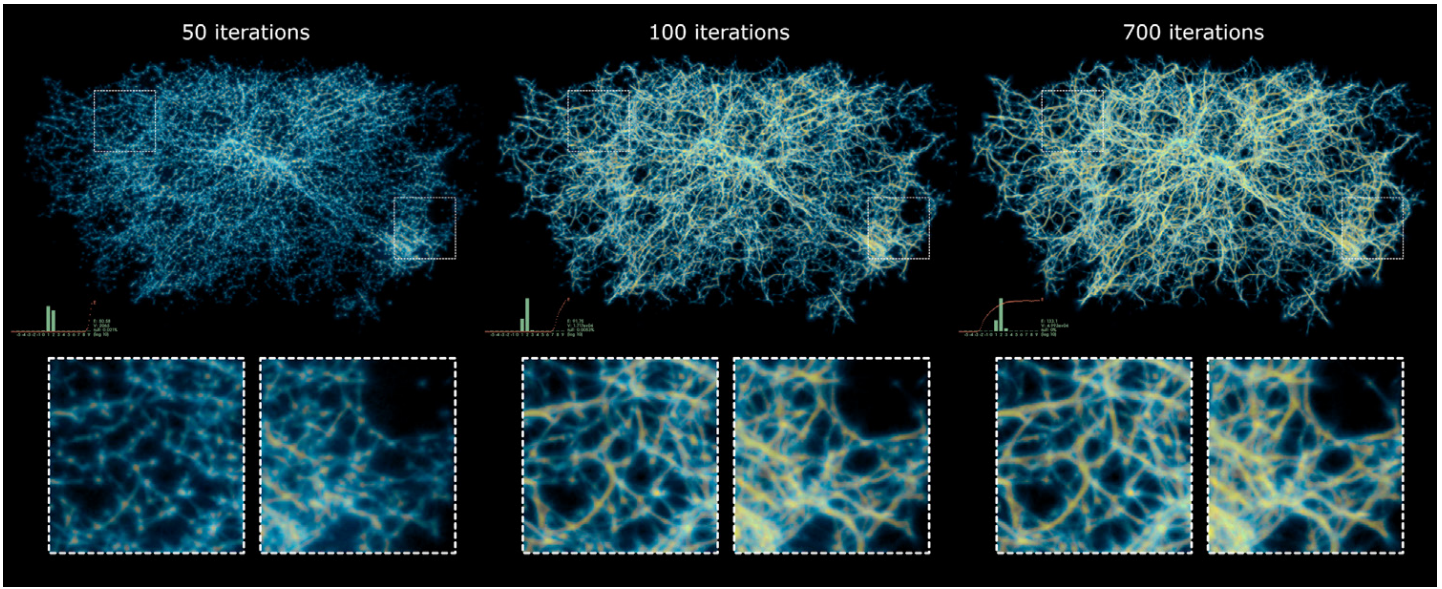

**Figure 21:** *Convergence behavior of MCPM when fitting to the 37.6k SDSS galaxies. The full fit takes about 700 MCPM iterations, but even after a fraction of the full fitting time we obtain a decent estimate of the eventual converged polyphorm (as seen in the insets below), allowing for early adjustments of MCPM's parameters.*

### 7.6 Distribution analysis

In analyzing the polyphorm patterns recovered by MCPM it is important to ask the question whether these are not arbitrary: just random, haphazard connections between neighboring points.

A detailed look into the input data and the resulting fits shows that this is not the case: as analyzed in Figure 22, many individual data points come together to give rise to any single filament. This matches the prior domain knowledge: the average distance between neighboring galaxies is in the order of 100 kpc, while the Cosmic web structures occur roughly between 1 and 100 Mpc. Crucially, it is the galaxies (or dark matter halos) whose distribution is aligned with the nodes of the Cosmic web, not the other way around. Their locations therefore sample the much *smoother high-level structure* of the Cosmic web, and this ultimately is what MCPM traces.

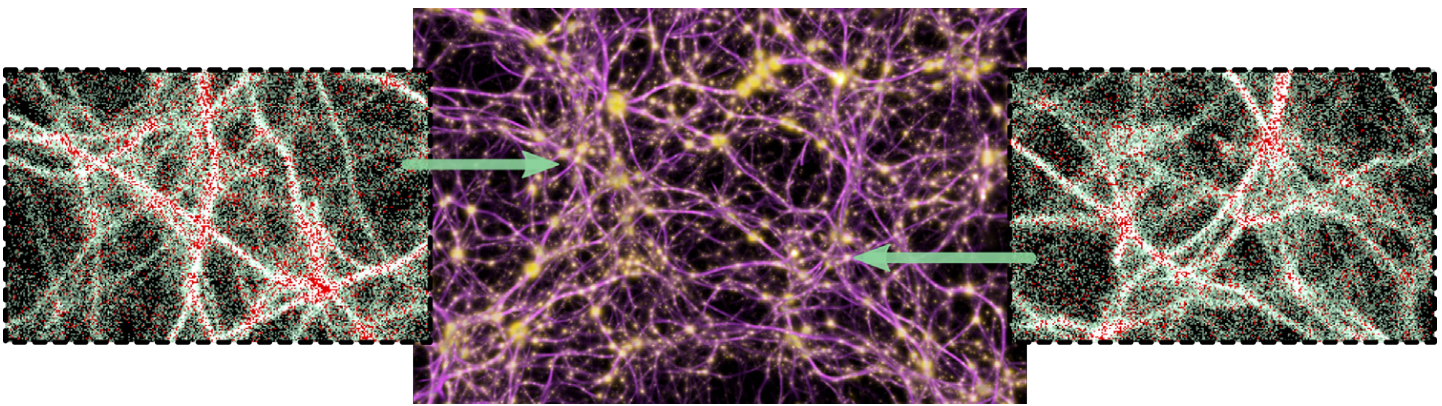

**Figure 22:** *Zooming in a section of the Bolshoi-Planck dataset about 50 Mpc wide (middle) and the filaments reconstructed therein. The magnifications show the raw data: dark matter halos as single red pixels and the MCPM agents in white. We see that the halos sample the domain with a significantly higher frequency than the actual reconstructed filaments.*

The above is of course subject to the configuration of the model. Our fits to both B-P and SDSS data use the sensing distance of 2.5 Mpc (as calibrated by fitting to the former). Structures significantly smaller than this would not be captured in the fit: they are essentially invisible to the agents. This is what makes the domain-specific knowledge valuable -- it constrains the fitting process and prevents us from obtaining nonexistent structures.

Going one step further, in Figure 23 we zoom in on *individual knots* (intersections) of the reconstructed filaments. Here the visualization segments the trace field into three bins: low (blue), medium (green) and high (red) density values. The agents' distribution here agrees with the astronomical intuition: the cores of the filaments and the knots receive the highest trace densities, which then fall off towards the outskirts of the observed structures.

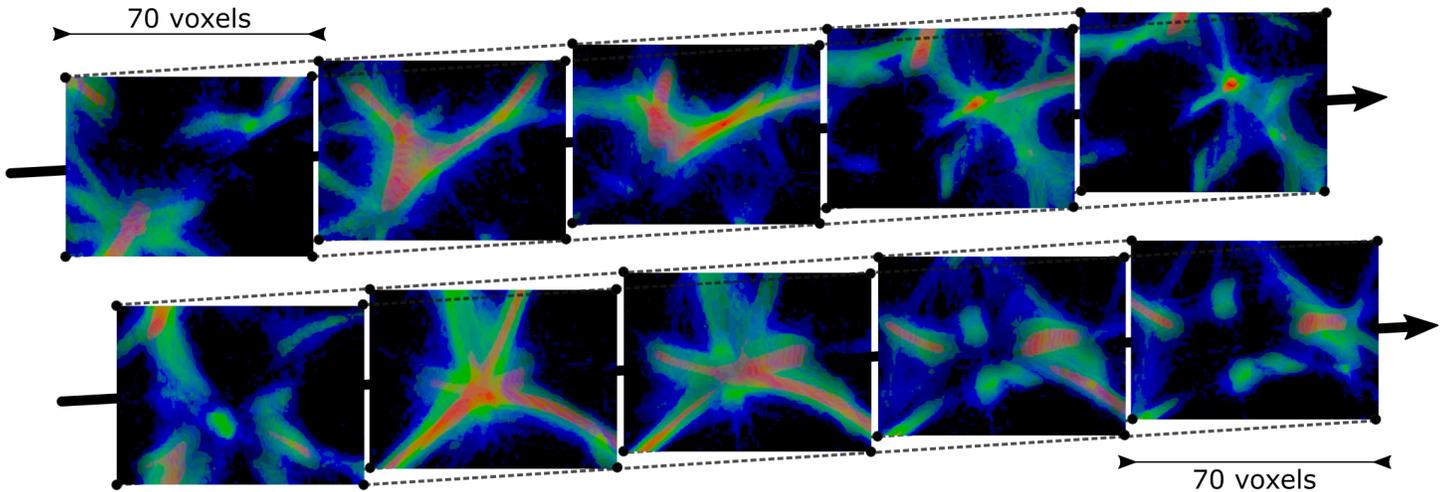

**Figure 23**: Two examples of complex knot specimens found in the Bolshoi-Planck data. These are regions about 10 Mpc wide, corresponding to 70x50x50 voxels (each visualized slice is 10 voxels thick). We observe the knots and filament cores to have the highest density (red), with a gradual fall-off outwards (green to blue).

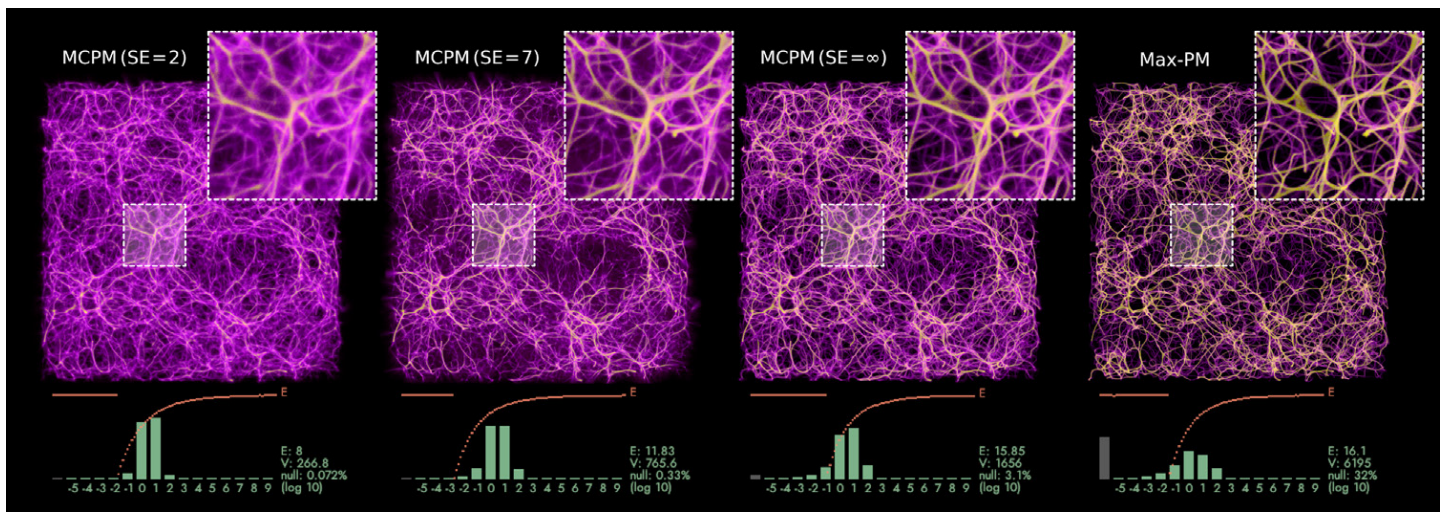

**Figure 24**: Fitting to Bolshoi-Planck data with variable Sampling exponent (SE). With increasing SE we obtain sharper polyphorms with reduced connectivity and worse data coverage (left to right). In the extreme case of Max-PM, the network condenses into nearly discrete filaments with a poor coverage of the input dataset (see the 'null' field in the plots, represented by the leftmost gray bar in the density histograms).

## 7.7 Mutation probability

The agents' decision whether to branch out, and with what probability, is one of the key determinants of the shape and complexity of the generated polyphorms. Here we explore this aspect of the model.

The variant of MCPM defined in Section 6 uses a probabilistic formulation for the mutation decision, controlled by a single scalar parameter: sampling exponent, SE for short. Increasing the value of SE leads to sharper and more concentrated filamentary structures (Figure 24), which also increases the energy of the fit. On the downside, the number of data points *not covered* by the fitted polyphorm increases, and for the limit case of infinite SE we miss more than 3% of data. The fitting process therefore involves finding a tradeoff between acuity of the structure and its coverage. In result, our fitting to the astronomical datasets uses moderate values of SE, usually between 2.2 and 4.

In the right panel of Figure 24 we compare to Max-PM [Jones2010]. Here we used 8 directional samples to compute the maximum of the deposit for the agents to follow in every step. As discussed in Section 5, this increases the structural acuity even further, optimizing the network by erasing smaller filaments, rounding off loops etc. While a sensible behavior for Physarum, this behavior causes a dramatic reduction of the network connectivity. As a result, 32% of input data points are missed by Max-PM, making this a poor fit (in spite of the seemingly high fitting energy $I$, cf. Section 7.3). Section 7.10 offers a more detailed quantitative comparison between MCPM and Max-PM.

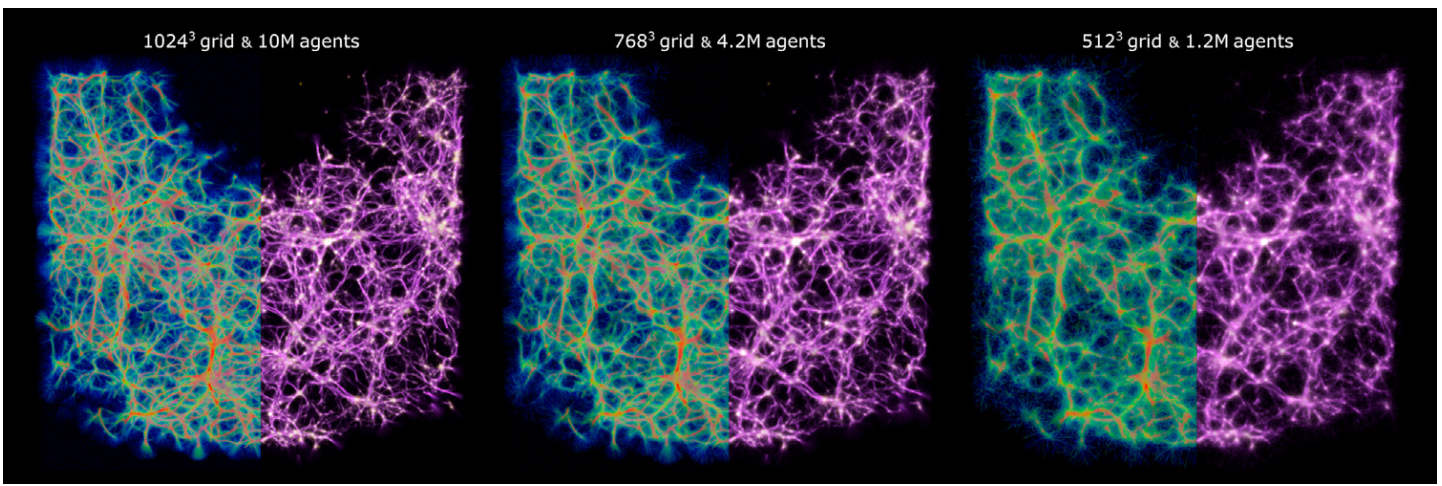

**Figure 25**: *A thin vertical slice of the Bolshoi-Planck data (about 15 Mpc in depth), showing fits with varying resolution of the trace and deposit fields. We also scale the number of agents in proportion to the number of voxels in the simulation grids. The left halves of each panel visualize segmented MCPM density, while the right halves overlay the filament structure (in purple) on top of data deposit (pale yellow halos).*

## 7.8 Trace & deposit field resolution

Here we examine the scaling of MCPM with respect to the resolution of the deposit and trace fields' representation.

In Figure 25 we compare three different field grid resolutions (with proportionally scaled numbers of agents), keeping the remaining model parameters fixed at the nominal values. We observe that while the amount of reproduced detail decreases with resolution -- and the amount of MC noise increases as the number of agents goes down -- the *overall structure* of the polyphorm remains largely intact. This is a desirable outcome, as we want to avoid a tight coupling between the precision of the representation and the geometry of the reconstructed network.

## 7.9 Dataset decimation

The ability of the model to retain structure is important when it is trained on dense data (such as the Bolshoi-Planck halos) and then transferred to incomplete data from observations.

In Figure 26 we look at the *structure retention* when reducing the number of dark matter halos in the BP dataset by thresholding the halo mass. By keeping only the heavier subset of halos we emulate the actual observations: the smaller, dimmer galaxies are less likely to be captured by SDSS and other such surveys. We see that even when decimating the input data by nearly an order of magnitude (from 870k halos down to about 100k), MCPM finds a good portion of the filaments -- all of the larger ones and some smaller ones. This gives us a clue about how much structure might be missing when fitting to the SDSS data. In Burchett et al. [2020] we explain the process of matching the density of these two distinct datasets in more detail.

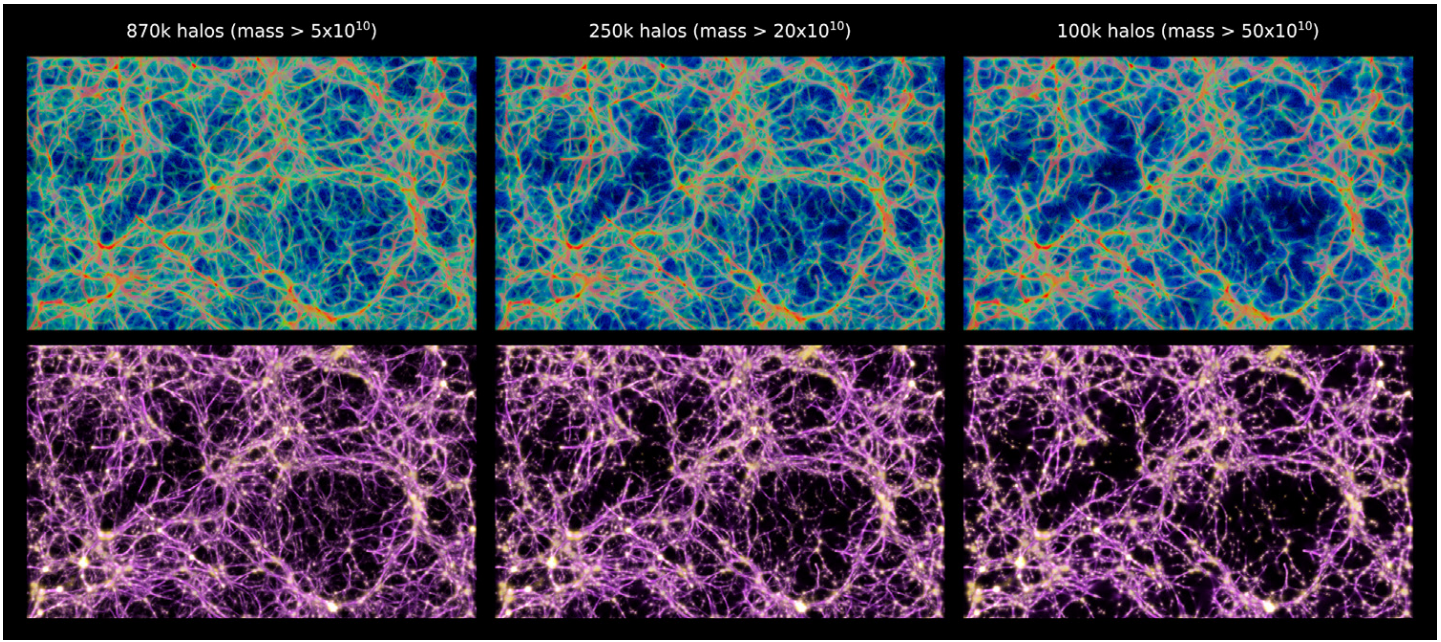

**Figure 26:** *Fitting to a variable number of Bolshoi-Planck halos, and otherwise using the standard parametrization for this dataset. While we observe a reduced number of detected filaments with fewer halos, the overall structure is preserved without major differences.*

## 7.10 Quantitative comparison between MCPM and Max-PM

We conclude this section with a detailed numerical study of MCPM in comparison to Max-PM [Jones2010], complementing the preceding qualitative results. In this experiment we fit both models to the Bolshoi-Planck simulation data (Section 7.4). Focusing on the accuracy of the fit, we do not apply any thresholding and retain all 12M dark matter halos, with the ultimate aim to match the reference density field produced by the simulation. In contrast to the halo catalog, the reference field represents the actual density of the simulated matter tracers, binned in a $1024^3$ grid.

To measure the extent to which MCPM and Max-PM match the reference data, we repeat the same procedure used to obtain the cosmological overdensity mapping in Burchett et al. [2020]. We first fit to the full 12M-halo catalog using both MCPM and Max-PM in a $1024^3$ simulation grid (which allows us to establish a one-to-one mapping to the reference grid without any interpolation). We then *standardize* both fits to make sure their means and variances match, thus obtaining comparable distributions. Finally, we compute 2D histograms binned over the fitted and the target density ranges, recording the mutual correlations between the corresponding voxels in 3D space.

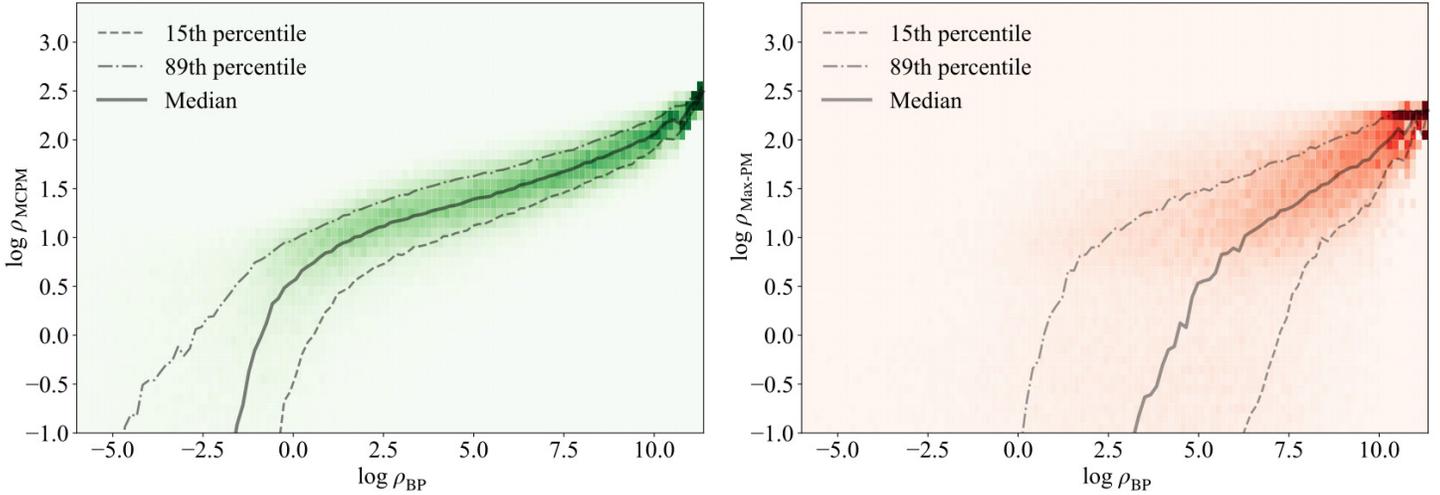

**Figure 27**: 2D histograms over the log-2 density ranges, comparing the fitted density fields (vertical axes -- MCPM in the left, Max-PM in the right) to the reference Bolshoi-Planck density field (horizontal axes). For every row of the histogram we plot the median of the respective conditional distribution and its approximate bounds (the chosen percentile values are asymmetric due to the slight skewness of the distributions).

The results are summarized in Figure 27. We observe two key differences between MCPM and Max-PM. First, MCPM covers a significantly wider range of the reference density field, roughly by 5 orders of magnitude in the low end of the distribution. In practice, this means that MCPM can capture much fainter features of the target structure, and characterize smaller substructure within the Cosmic web. This mirrors our findings from Section 7.7, where we see a significant portion of the halos being missed by the Max-PM fit. Second, the MCPM fit has considerably better precision compared to Max-PM. This is apparent from the spread of the histograms, which we visualize by plotting the percentile bounds of the per-row conditional distributions.

From these results, we conclude that MCPM is able to reconstruct an accurate proxy of the reference density field from very sparse data, i.e., using 12M weighted halo points compared to the 1B voxels in the reference density grid. Compared to Max-PM, MCPM improves the Cosmic web density reconstruction by several orders of magnitude, which in practical terms means extending the reconstruction from only the dense structures (knots and major filaments) all the way to the edges of cosmic voids.

## 8. Discussion

The presented MCPM model is suitable for revealing geometric structures in incomplete data, as demonstrated by mapping the Cosmic web filaments in SDSS galaxy data. It is especially effective if a training dataset is available to determine the prior parameters (as we did with fitting to Bolshoi-Planck dark matter halos) and if expert knowledge is available to determine the appropriate probability distributions (as we did by using the Maxwell-Boltzmann distribution for agent distance sampling).

Fitting MCPM to data is a robust, reliable process. For any given parametrization the model eventually converges to the same outcome (minus the MC variance), regardless of the initial distribution of the agents. To our surprise, we have not observed any tendency of the model to get stuck in local optima.

We now conclude the paper with an open-ended discussion about the current state of the model and what research directions we plan to explore in the future.

## 8.1 Mapping the Cosmic web

In hindsight, using a virtual physarum machine to reconstruct the Cosmic web seems like a very intuitive thing to do: MCPM has already provided a viable solution for several astronomical use cases [Burchett2020, Simha2020]. We continue to inquire into the implications of this possibility. This includes:

- further ways to validate the model's performance in the presented tasks, as well as in new contexts;
- continue the design pathway towards a more specific (accurate) variant of MCPM tailored for astronomy;
- developing deeper mathematical foundations for these models;
- topological analysis for graph extraction (directly applicable to catalog the Cosmic web filaments, but also in other tasks; see below).

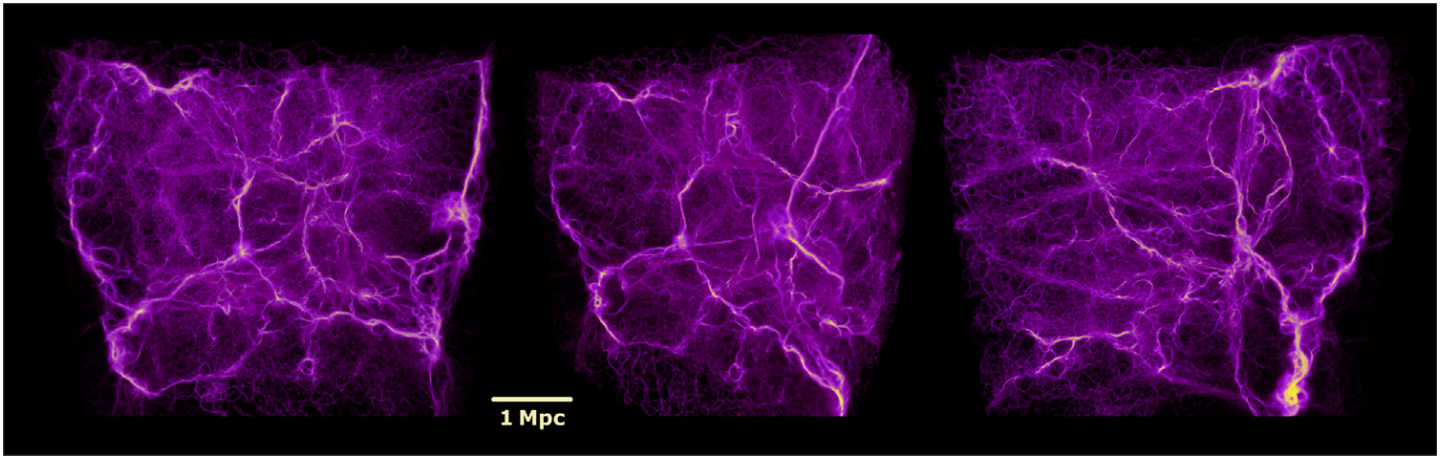

**Figure 28**: *Experimental fit to a single page of the EAGLE dataset (about 1M data points representing quanta of gas, selected as the heaviest slice from the original 16M particles). MCPM here serves both as a reconstruction and a visualization method, providing a convenient estimate of the gas density.*

This applies also to using MCPM for examining and quantifying *other astronomical datasets*. One example is the EAGLE dataset [Schaye2015], which includes billions of simulated macro-particles (tracers) of the intergalactic medium but over a smaller high-resolution volume and including gas and the resulting complex hydrodynamical astrophysic. For a demonstration, in Figure 28 we fit to a single page (a sub-volume of a single redshift snapshot) of these data: a cubic region with roughly 4 Mpc across, i.e., two orders of magnitude smaller than the Bolshoi-Planck data. This is the smallest scale where the Cosmic web starts becoming apparent -- on the edge of circumgalactic halos, which are well resolved in these data. The structures that MCPM finds here are intriguing, and further investigation involving observational data will show whether they are real and significant.

## 8.2 Temporally varying data

In this paper we have focused on static input data, and sought an equilibrium solution that best represents it. Of course, Physarum itself does not behave this way -- it never truly halts its growth and remains in a constant state of flux during its plasmodium stage. This behavior happens in response to environmental changes around the organism.

For MCPM, the above translates to handling *dynamic, time-varying data*. Cosmological inquiry routinely uses time-varying data, since the evolution of the universe can be approximately traced by matter halos (Section 7.4) that vary in position, mass and numbers. Beyond cosmology, such data are produced by models of population dynamics, traffic, epidemiology, neuronal networks and others.

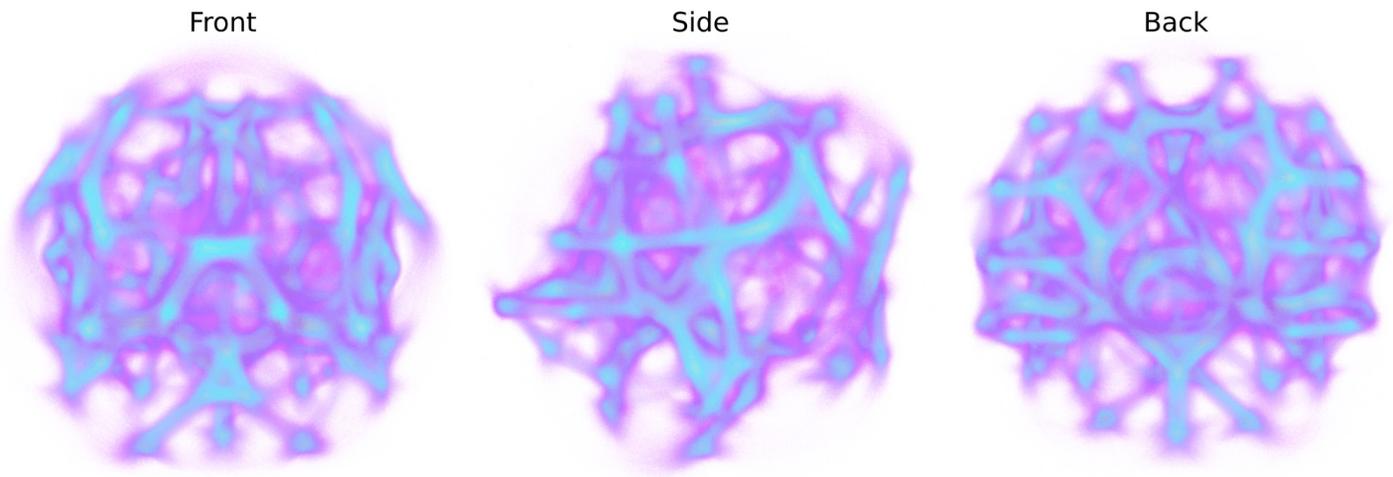

**Figure 29**: MCPM reconstruction of the brain connectome from a sparse representation (105 high-level nodes weighted by their mutual connectivity). The symmetry (or lack thereof) in the connectivity patterns is easily distinguishable, as well as the localization of the information 'flow'.

As an illustrative example, we consider the problem of 3D edge bundling in complex network visualization [Holten2009, Bach2017, Ferreira2018]. We used MCPM to fit and visualize a human brain connectome [Xu2019], a set of 105 high-level nodes representing different sub-regions of the brain (so-called *modes*). To simplify this experiment (Figure 29), we disregard the connectome edges and simply weigh each node by its overall connectivity strength. The result is a fuzzy activity map. In future, adding the edges as a directional guide for the agents could provide us with a more intricate spatial map with hierarchical grouping of connections and color differentiation between the modes.

Currently, MCPM can be used for time-varying data if the temporal evolution of the dataset is matched with the model dynamics. For dynamic data with stored snapshots -- such as the cosmological simulations -- this can be achieved by fitting to a weighted combination of two adjacent snapshots, similar to *keyframing* in computer animation. In the more general case of dynamically evolving data (e.g., produced in real-time by a simulation), the sensing decisions of the agents would need to incorporate additional knowledge about the data, for instance by extrapolating the current state of the dataset forward in time.

## 8.3 Biomimetic modeling

Whether the polyphorm is a natural 3D generalization of the real-world Physarum's plasmodium is debatable. But undeniably the patterns generated by MCPM do have an organic quality to them.

For instance, networks generated on low-discrepancy sampled point sets -- even with spatially varying point density -- resemble *organic scaffoldings* found in diatoms, corals, and bone marrow Figure 30. This could have applications in tissue engineering [Derby2012, Thrivikraman2019], either via 3D printing or other fabrication methods.

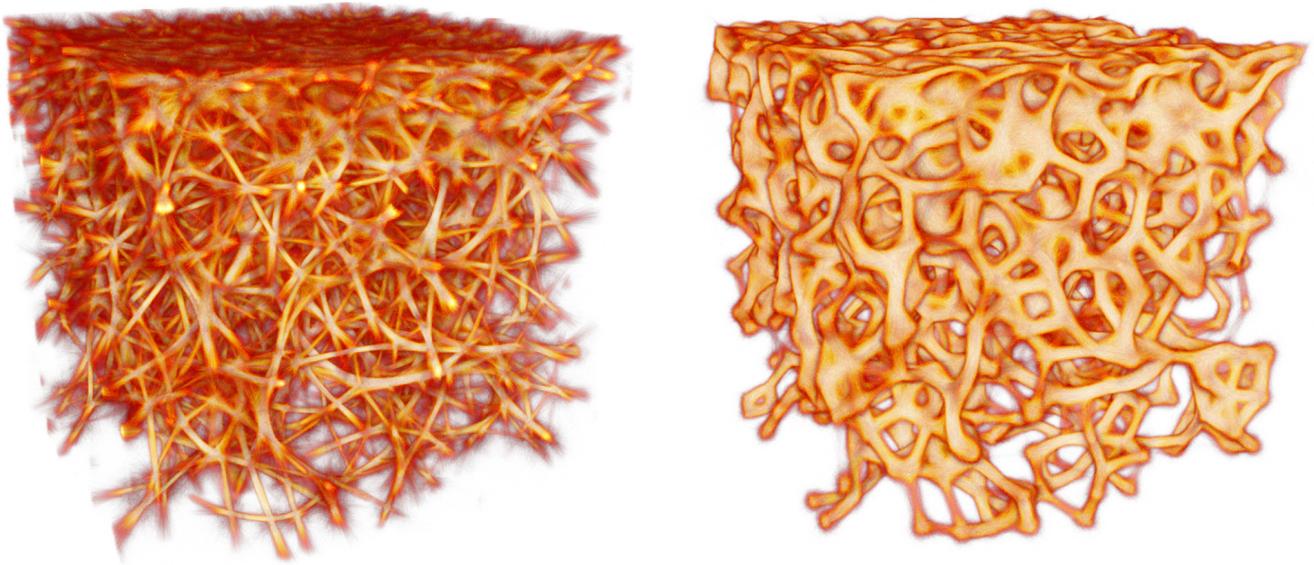

**_Figure 30_**: _Polyphorms generated on 4k low-discrepancy points with vertically biased sampling density. By varying MCPM's geometric parameters of the model we obtain structures with different volumes and connectivity._

3D printing in general would benefit from organically inspired internal object scaffoldings, since the predominantly used periodic lattices do not lead to even stress distribution [Safonov2017]. MCPM could be applied in this context to generate aperiodic quasi-crystalline structures that mimic scaffoldings produced by topological optimization methods. Since MCPM is designed with the intent of being calibrated on reference data, we could find structurally sound parametrizations by fitting to topologically optimized scaffoldings, and then apply these efficiently during the preprocessing of each 3D print with sensitivity to their peculiar geometries.

Beside scaffoldings, it would be very interesting to see whether other natural networks -- like mycelium or even neuronal networks -- could be modeled by specialized variants of MCPM. And similarly to the real organism, adding a repulsive gradient could lead to further and more expressive types of polyphorm geometries.

## 8.4 Artificial networks

Many kinds of constructed (abstract) networks are conveniently represented as graphs (social, logistic, dependency, decision, automaton, and other such networks). Some network optimization problems are also amenable to Physarum-based solutions [Sun2017]. Especially problems with a lot of inherent uncertainty like traffic routing and disease spreading could benefit from MCPM, thanks to its customizable probabilistic form.

In Zhou et al. [2020] we investigate the application of MCPM to _language embedding_ data. Language embeddings [Mikolov2013, Vaswani2017] are compressed high-dimensional representations produced by transformer neural nets. Their usefulness and predictive power is unmatched in natural language processing, as demonstrated by the recent GPT-3 model for instance [Brown2020]. Our work [Zhou2020] presents early evidence that embeddings could have an internal structure described by an optimal transport network, see the example in Figure 31.

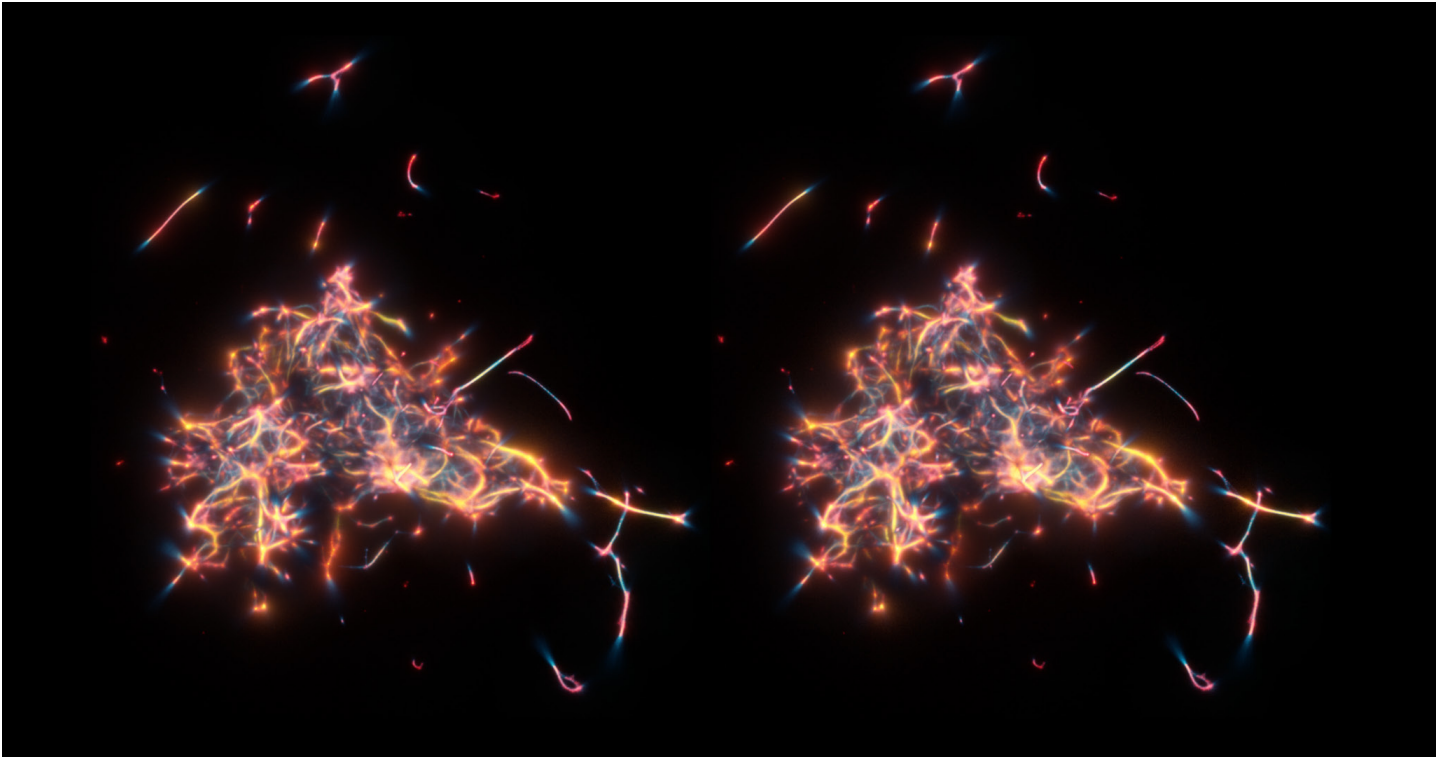

**Figure 31:** *Auto-stereoscopic rendering of the MCPM fit to the Word2Vec language embedding data. All 300k word tokens contained in Word2Vec (representing the Wikipedia 2017 corpus) are projected from the original 512-dimensional embedding space down to 3D using UMAP (red clusters), and then reconstructed with MCPM (yellow-cyan colormap). The result reveals a complex polyphorm interconnecting the main central cluster, as well as a number of smaller, isolated clusters. The 3D structure can be seen by crossing eyes until the two sides merge.*

Given that MCPM can produce robust continuous networks up to three dimensions, we see a great potential in generalizing it to **higher dimensions**, which now becomes feasible thanks to the stochastic formulation and binary sampling stencil. Visualizing and automatically navigating high-dimensional latent spaces such as the language (and other kinds of) embeddings would be very beneficial, as insights into these increasingly popular compressed representations are still lacking.

## 8.5 Structure-sensitive learning

MCPM bears close resemblance to other self-organizing methods for *unsupervised learning*. The self-organizing map [Kohonen1982] and neural gas [Martinetz1991, Fritzke1994] are good examples of such methods.

The self-organizing map is a topology-preserving dimensionality reduction algorithm: similar to UMAP, it projects high-dimensional training data to a low-dimensional space (typically 2D) while preserving neighborhood relationships. During the training process the map reorganizes itself by locally 'gravitating' towards clusters of data in the original high-dimensional space. The neural gas network operates similarly; however, its topology is not predetermined -- instead, neural gas assumes the presence of a high-dimensional topological structure which it tries to adapt to.

Compared to these methods, MCPM does not maintain explicit connectivity between the agents; instead, they flow freely in the domain and all information exchange is done through the deposit field. As a result, the topology of the fit is implicit in the continuous trace density estimate. Recovering the topology explicitly, e.g. as a graph would be valuable both in the context of the Cosmic web reconstruction (to provide a classification mechanism for the filaments and knots) but also outside of it; adapting the neural gas algorithm to work on density fields seems like a viable option here.

The advantage MCPM offers over these methods is the continuous nature of the reconstructed network, permitting its interpretation as a density field [Burchett2020, Simha2020] or a throughput field [Zhou2020]. This is enabled by the core design of MCPM: the solution is not a fixed point in the search space, but rather, an average of the *possible dynamic agent trajectories*.

### 8.6 MCPM as a complex dynamic system

A noteworthy aspect of the model is that we often find the best-fitting polyphorms near *tipping points* of its dynamic behavior. This happens especially with regard to changing the scale-related parameters like the Sensing distance (SD). In a typical scenario we would start fitting with high values of SD, then gradually decrease it. That in turn decreases the size of reconstructed features, up to the point where the system reaches a supercritical state and 'dissolves' in Monte Carlo noise. That *tipping point* would usually be a local energy minimum. It would be interesting to further examine this aspect of the model more formally.

Based on this, we speculate that MCPM could detect its tipping points, e.g., through a redesign of the loss function (Section 7.3). This could lead to a specialized *physarum-based AI*. Such an enhanced model would be able to locally modify its parametrization in order to fit to heterogeneous data. One possible path towards that would be to implement a spatial voting scheme where the agents would locally contribute information from their recent exploration of the domain and contribute their individual parameter estimates.

## 9. Conclusion

We have presented Monte Carlo Physarum Machine (MCPM), a stochastic computational model inspired by the growth and foraging behavior of Physarum polycephalum 'slime mold'. MCPM takes as its input a weighted point cloud in 2D or 3D, and produces a continuous transport network spanning the point data as an output. The output is represented by a scalar density field expressing the equilibrated spatial agent concentration, and we refer to this entity as a **polyphorm** to capture both its geometric quality as well as origin in a single term.

The main contribution of this work with regard to its predecessor [Jones2010] is the **probabilistic approach** for modeling the agents' behavior. This makes the model configurable for a range of applications. The specific version of MCPM analyzed in this paper has been designed for the application in astronomy, aiming to make sense of large cosmological datasets and reconstruct the filamentary structure of the Cosmic web. Thanks to MCPM, we have made new discoveries relating to the structure of the intergalactic medium, which composes the Cosmic web [Burchett2020, Simha2020]. We will continue this line of inquiry and seek out other applications where MCPM can be of use, such as scientific visualization [Elek2021] and natural language processing [Zhou2020].

Our **open-source implementation** of MCPM is called *Polyphorm* (`github.com/CreativeCodingLab/Polyphorm`). *Polyphorm* runs primarily on the GPU, taking advantage of the inherent parallelizability of MCPM, and provides an interactive reconstruction and visualization experience. We invite the reader to use the software, import their own domain-specific data and enrich our understanding of where such a model could be meaningfully applied.

There seems to be a degree of **universality** to MCPM -- our experience so far has been that people with very different backgrounds have been able to identify with the generated polyphorm geometries in their specific ways. This is not surprising, since the agents approximately follow paths of least resistance. This energy-minimization property will likely be part of many natural or abstract systems.

We believe that this work reinforces the possibility of Nature as an information processing entity -- a computable Universe. The piece of evidence we contribute is that the small set of mathematically coherent rules presented here might apply to such different phenomena at vastly divergent scales.

# Acknowledgements

The authors wish to thank Jan Ivanecký for developing an early prototype of the *Polyphorm* software and valuable technical assistance during the development, Daisuke Nagai for insightful advice, and the anonymous reviewers whose feedback strengthened the paper. We gratefully acknowledge the NVidia Hardware Grant for the donation of the TitanX development GPU.